\documentclass[11pt]{article}
\usepackage{cite}
\usepackage{array}
\usepackage{epsfig}
\usepackage{hyperref}
\usepackage{amsmath}
\usepackage{multirow}
\usepackage{amssymb}
\usepackage{color}
\usepackage{indentfirst}
\usepackage{pstricks}
\numberwithin{equation}{section}
\def\rmd{{\rm d}}

\newcommand{\el}[1]{\label{#1}}
\newcommand{\er}[1]{\eqref{#1}}
\newcommand{\df}[1]{\textbf{#1}}

\newcommand{\lb}{\left(}
\newcommand{\rb}{\right)}

\newcommand{\nn}{\nonumber \\}
\newcommand{\p}{\partial}

\newcommand{\cd}{\nabla}

\newcommand{\ba}{\begin{eqnarray}}
\newcommand{\ea}{\end{eqnarray}}
\newcommand{\be}{\begin{equation}}
\newcommand{\ee}{\end{equation}}
\newcommand{\bay}[1]{\left(\begin{array}{#1}}
\newcommand{\eay}{\end{array}\right)}
\newcommand{\eg}{\textit{e.g.} }

\newcommand{\zt}[1]{\textrm{#1}}

\newcommand{\Tr}{\mbox{Tr}}
\def\xa{{\alpha}}

\def\xD{{\Delta}}
\def\xe{{\epsilon}}

\def\xg{{\gamma}}
\def\xG{{\Gamma}}
\def\xk{{\kappa}}

\def\xl{{\lambda}}

\def\xo{{\omega}}

\def\xs{{\sigma}}
\def\xS{{\Sigma}}
\def\xt{{\theta}}

\def\xT{{\Theta}}

\def\CD{{\cal D}}

\def\CL{{\cal L}}
\def\CM{{\cal M}}
\def\CN{{\cal N}}

\def\BC{\mathbb{C}}

\def\BN{\mathbb{N}}

\def\BZ{\mathbb{Z}}

\begin{document}

\baselineskip18pt

\begin{titlepage}

\vspace*{0.7cm}

\centerline{\Large \bf 3d SCFT on Conic Space}
\vskip0.45cm
\centerline{\Large \bf as Hologram of}
\vskip0.45cm
\centerline{\Large \bf Charged Topological Black Hole}
\vspace*{1.3cm}
\centerline{\Large Xing Huang $^a$, \ \ \ Soo-Jong Rey $^{a,b}$, \ \ \ Yang Zhou $^b$}
\vspace*{1.3cm}
\centerline{\sl School of Physics \& Center for Theoretical Physics}
\vspace*{0.15cm}
\centerline{\sl Seoul National University, Seoul 151-747 \rm KOREA $^a$}
\vspace*{0.15cm}
\centerline{\sl Center for Quantum Spacetime, Sogang University, Seoul 121-742 \rm KOREA $^b$}
\vspace{2cm}
\centerline{ABSTRACT}
\vskip0.75cm
\noindent
We construct three-dimensional ${\cal N}=2$ supersymmetric field theories on conic spaces. Built upon the fact that the partition function depends solely on the Reeb vector of the Killing vector, we propose that holographic dual of these theories are four-dimensional, supersymmetric charged topological black holes. With the supersymmetry localization technique, we study conserved supercharges, free energy, and R\'enyi entropy.
At planar large $N$ limit, we demonstrate perfect agreement between the superconformal field theories and the supersymmetric charged topological black holes.

\end{titlepage}

\section{Introduction}
In supersymmetric field theories with conserved R-charges, the localization technique, pioneered by Pestun \cite{Pestun:2007rz} in four dimensions, Kapustin, Willett and Yaakov \cite{Kapustin:2009kz} in three dimensions, and Kallen and Zabzine \cite{Kallen:2012cs} in five dimensions, provides exact computation of a certain class of physical observables. Once combined with physical considerations, these results provide useful data sets for understanding nonperturbative dynamics. Essential prerequisites to these studies is the construction of supersymmetric field theories on curved backgrounds. It is now understood quite extensively how to put supersymmetric field theories on Riemannian manifolds.

The purpose of this paper is to refine previous investigations to the supersymmetric field theories on singular space and its holographic dual. Because of singularities, the base spacetime is no longer a manifold and the previous constructions may fall short of its validity. Nevertheless, in this paper, we shall show that the localization technique can be made to work even on singular spaces provided the theory under consideration is conformally invariant. For concreteness of our investigation, by singular space, we specifically refer to the branched sphere $\mathbb{S}^3_q$ that is formed from sphere by inserting conic singularities, where $(q-1)$ is a deformation parameter away from the round three-sphere $\mathbb{S}^3$.

The idea is that, firstly, $\mathbb{S}^3_q$ has the same Reeb vector as the ellipsoid $\widetilde{\mathbb S}^3_b$ and that, secondly,  this space is conformally equivalent to $\mathbb{S}^1 \times \mathbb{H}^2$, at least, locally. The first point implies that, for ${\cal N}=2$ supersymmetric field theories,  the partition function on $\mathbb{S}^3_q$ is the same as the partition function on $\widetilde{\mathbb{S}}^3_b$. The second point implies that, if the theories are superconformal invariant, the partition function on $\mathbb{S}^3_q$ is the same as the partition function on $\mathbb{S}^1 \times \mathbb{H}^2$. These chain of equivalences also hold for other observables  than identity operator so long as they are in the orbit of the conserved supersymmetries on $\mathbb{S}^3_q$.

The above idea also suggests that, in the large $N$ limit, these SCFTs on the branched sphere $\mathbb{S}^3_q$ are holographically dual to a topological black hole (TBH) in four-dimensional anti-de Sitter spacetime, whose horizon takes the shape of $\mathbb{H}^2$. We shall refer this proposal as the TBH / qSCFT correspondence. The black hole is charged, which reflects the fact that conical singularity  of $\mathbb{S}^3_q$ is accompanied by a background vector field dual to the conserved R-symmetry. The black hole is extremal, which reflects the fact that the SCFTs on $\mathbb{S}^3_q$ preserve two supercharges. To test our proposed TBH / qSCFT correspondence, we computed the free energy and R\'enyi entropy of the TBHs. The results show perfect agreement with the exact results of the qSCFT in the large $N$ limit.

This paper is organized as follows. In section \ref{sec:qSCFT}, we formulate the qSCFTs -- the ${\cal N}=2$ SCFTs on various branched spheres, including $\mathbb{S}^3_q$. We first analyze the charged Killing spinor equations in the background off-shell supergravity fields. We then compute the partition functions on different branched spheres by localization and show that their form remains exactly the same as the ellipsoidal three sphere $\widetilde{\mathbb{S}}^3_b$. We analyze the Reeb vectors on these spaces and show that, partition functions only depend on Reeb vectors on branched spheres.
Built upon these observations, we extract the partition function of qSCFTs and show that its free energy $F(q)$ in the large $N$ limit takes the form
\be\label{q_free_energy}
F(q) = {1 \over 4} \left( \sqrt{q} + {1 \over \sqrt{q}} \right)^2 {F}(1)\ .
\ee
We also extract the R\'enyi entropy and show that it takes the form
\be\label{q_renyi}
S(q) = {3q + 1 \over 4q} S(1), \quad \mbox{and hence} \quad S(\infty) = {3 \over 4} S(1)\ .
\ee
In section \ref{sec:TBH}, we study the charged topological black hole solution in the context of four dimensional $\cal N$$=2$ gauged supergravity. We first analyze the 4d Killing spinor equation on this background and show that the integrability condition determines the BPS condition for the black hole, namely the mass-charge relation. We then discuss two supersymmetric black hole solutions, neutral massless and charged BPS black holes. In section \ref{sec:TBH/qSCFT}, we show the TBH$_4$/qSCFT$_3$ correspondence. We first fix the charged topological black hole solution by matching temperature and chemical potential to the boundary field theory on $\mathbb{S}^1\times\mathbb{H}^2$ and compute the free energy and the R\'enyi entropy, which precisely agree with the localization results (\ref{q_free_energy})(\ref{q_renyi}) of qSCFT$_3$ in the large $N$ limit. We then analyze the supersymmetry of this TBH$_4$ and show that it is BPS. We further find the Killing spinor solutions for this TBH$_4$ and show that it preserves the same number of supercharges as the boundary field theory. We conclude and discuss future questions in section \ref{sec:conclusion}.


\section{qSCFT$_3$}
\label{sec:qSCFT}
Supersymmetric field theories were constructed on round three-sphere \cite{Kapustin:2009kz}, ellipsoid and squashed sphere \cite{Hama:2011ea} \cite{Imamura:2011wg}. The partition function on $\mathbb{S}^3$ was found not to depend on the size of $\mathbb{S}^3$, and this is a consequence of the conformal fixed point the theory flows to. The partition function on $\widetilde{\mathbb{S}}^3_b$ was found to depend on squashing parameters of $\widetilde{\mathbb{S}}^3_b$. A refinement of such construction is supersymmetric field theories on a three-sphere with conical singularities \cite{Nishioka:2013haa}. The conical singularity is specified by a parameter $q \in \mathbb{R}$. We can think of $\mathbb{S}^3_q$ as a $q$ deformation of round 3-sphere and $\widetilde{\mathbb{S}}^3_q$ as a $q$ deformation of squashed sphere or ellipsoid.

In this section, we shall construct supersymmetric field theories on $\mathbb{S}^3_q$ and $\widetilde{\mathbb{S}}^3_q$, following the systematical approach~\cite{Dumitrescu:2012ha, Closset:2012ru, Closset:2012vp, Closset:2012vg}, which was initiated in~\cite{Festuccia:2011ws}. 
The construction is based on the rigid limit of three dimensional supergravity that couples to the R-multiplet of the field theory. We are particularly interested in three dimensional $\cal N$ = $2$ theories with a $U(1)_R$ symmetry. Note that such construction is equivalent to the construction of superconformal field theories (SCFTs) on curved manifold, because it is now known \cite{Klare:2012gn, Cassani:2012ri, Hristov:2013spa} that the solutions to the conformal Killing spinor equations are closely related to the solutions of the Killing spinor equations we will solve in this section \footnote{See also \cite{Closset:2012ru} for exemplification of this in Euclidean three-manifolds.}. Therefore, in what follows, we shall not distinguish the two constructions. The Poincar\'e supersymmetry algebra involving supercharges $Q_\alpha, \widetilde Q_\alpha$ of R-charges $\pm 1$ reads
\ba
\{Q_\alpha, \widetilde Q_\beta\} = 2\gamma^\mu_{\alpha\beta}P_\mu + 2 i \varepsilon_{\alpha\beta} Z\ ,\nn
\{Q_\alpha, Q_\beta\} =0\ ,\quad \{\widetilde Q_\alpha, \widetilde Q_\beta\} =0\ .
\ea
The supergravity multiplet contains the metric $g_{\mu\nu}$, two gravitini $\psi_\mu,\widetilde\psi_\mu$, an Abelian two-form connection $B_{\mu\nu}$ and two Abelian one-form connections $A_\mu$ and $C_\mu$. For auxiliary fields $B_{\mu\nu}$ and $C_\mu$, the dual of their field strengths are denoted by
\be\el{strengths_V_H}
V_\mu = - i \varepsilon_\mu^{\,~\nu\rho}\partial_\nu C_\rho\ ,\quad H = {i\over 2} \varepsilon^{\mu\nu\rho}\partial_\mu B_{\nu\rho}\ .
\ee
The metric $g_{\mu\nu}$ couples to the energy momentum tensor, the gauge connection $A_\mu$ couples to the $U(1)_R$ current, and $C_\mu$ couples to the central current whose charge is the charge $Z$ that appears in the supersymmetry algebra above. A given configuration of the background fields $A_\mu, V_\mu, H$ preserves supersymmetry if and only if the variations (parameterized by some choice of $\zeta$ and $\widetilde{\zeta}$) of the two gravitini fields vanish:
\be
\delta \psi_\mu = 0\ ,\quad \delta\widetilde\psi_\mu = 0\ .
\ee
In Euclidean signature, $\zeta$ and $\widetilde{\zeta}$ are independent complex spinors. These conditions essentially give the Killing spinor equation
 \be
\left(\nabla_{\mu}-iA_{\mu}\right)\zeta=-\frac{1}{2}H\gamma_{\mu}\zeta-iV_{\mu}\zeta-\frac{1}{2}\varepsilon_{\mu\nu\rho}V^{\nu}\gamma^{\rho}\zeta\ ,\label{eq:Killing_Spinor_Equation}
\ee where a solution $\zeta$ of it corresponds to a supercharge $\delta_\zeta$ taking R-charge $+1$, while a supercharge $\delta_{\widetilde\zeta}$ of R-charge $-1$ corresponds to a solution of
\begin{equation}
\left(\nabla_{\mu}+iA_{\mu}\right)\widetilde\zeta=-\frac{1}{2}H\gamma_{\mu}\widetilde\zeta+iV_{\mu}\widetilde\zeta+\frac{1}{2}\varepsilon_{\mu\nu\rho}V^{\nu}\gamma^{\rho}\widetilde\zeta\ .\label{eq:Killing_Spinor_Equation2}
\end{equation}
These equations generally tell us what background fields on three-dimensional space ${\cal M}_3$ allow a set of rigid supersymmetries.

\subsection{Killing spinors on Branched Spheres}
\label{subsec:Killing}
We would like to solve the Killing spinor equations for branched 3-spheres. These equations were analyzed in \cite{Closset:2012ru} \cite{Klare:2012gn}, where the three dimensional space ${\cal M}_3$ is a Riemannian three-manifold. There, it was shown that three-dimensional rigid supersymmetry requires an almost contact structure on ${\cal M}_3$, much the same way four-dimensional rigid supersymmetry requires an almost Hermitian structure~\cite{Festuccia:2011ws}. Here, we refine these results to  three-dimensional spaces containing conical singularities.

\subsubsection{$\mathbb{S}^3_q$}\label{S3q_kse}
This class of branched 3-spheres can be characterized by deformation of round 3-sphere. A quick way to see this is by dilating the metric while keeping domains of coordinates intact. The metric of the 3-sphere then turns into
\be
\label{3sphere}
\rmd s^2 = \ell^2 \left( \rmd\theta^2 +  \cos^2\theta \rmd\phi^2 + q^2\sin^2\theta \rmd\tau^2\right) \ ,
\ee
where, as said, the domains of $\theta, \tau, \phi$ are
\be
 \theta\in[0,\pi/2]\ ,\quad \tau\in [0,2\pi)\ , \quad \phi\in [0,2\pi)\ .
 \ee
If $q \neq 1$, the space has a conical singularity at the point $\theta=0$, otherwise regular everywhere else. We can regard the branched sphere as a deviation from the round 3-sphere 
parameterized by $q-1$. Therefore we expect that Killing spinor equations have minimal deviations from those for round sphere, with an additional background gauge field $A_\mu$. Thus we have $A_\mu\neq 0$ (modulo flat connection), $H\neq 0$ and $V=0$ and Killing spinor equations become a special case of (\ref{eq:Killing_Spinor_Equation}) and (\ref{eq:Killing_Spinor_Equation2})
 \ba
\left(\nabla_{\mu}-iA_{\mu}\right)\zeta &=&-\frac{1}{2}H\gamma_{\mu}\zeta\ ,\label{kseq1}\\
\left(\nabla_{\mu}+iA_{\mu}\right)\widetilde\zeta &=&-\frac{1}{2}H\gamma_{\mu}\widetilde\zeta\label{kseq2}\ .
\ea
Spinor covariant derivative is defined as
\be
\nabla_\mu \zeta = \partial_\mu\zeta +{1\over 4}\omega_\mu^{\,~ij}\sigma_{ij}\zeta\ ,
\ee where $\sigma_{ij}:={1\over 2}[\sigma_i, \sigma_j]$ and the spin connection $\omega_\mu^{\,~ij}$ is given in terms of the Christoffel connection $\Gamma_{\sigma \mu}^\nu$ by
\be\label{spinconnection}
\omega_\mu^{\,~ij} = e_\nu^i\partial_\mu e^{\nu j} + e^i_\nu e^{\sigma j}\Gamma_{\sigma \mu}^\nu\ .
\ee
To solve these equations, we use the fact that the round 3-sphere is the $SU(2)$ group manifold with group element $g$. The metric of the $SU(2)$ group manifold reads
\be
\rmd s^2 = \ell^2 \mu^m\mu^m = \ell^2\widetilde\mu^m\widetilde\mu^m\ ,
\ee where $m=1, 2, 3$, $\mu :=g^{-1}\rmd g$ and $\widetilde\mu :=\rmd g g^{-1}$ are left-invariant and right-invariant 1-forms, respectively. In  the left-invariant frame, the vielbeins are given by
\be
e^1 = \ell \mu^1\ ,\qquad e^2 = \ell \mu^2\ , \qquad e^3 = \ell \mu^3\ .
\ee
Likewise, the $q$-branched sphere can be constructed by rescaling $\rmd \tau$ in the vielbein to $q\,\rmd\tau$. We collected the vielbein and spin connection in Appendix~\ref{vielbeinS3q}. With the convention of the three-dimensional gamma matrices in terms of Pauli matrices as
\be\label{gammas}
\gamma_{1} = \sigma_1\ ,\qquad \gamma_{2} = \sigma_2\ , \qquad \gamma_{3} = \sigma_3\ ,
\ee
the spin connection $\omega_\theta^{\,~ij}\sigma_{ij}$ is proportional to $\gamma_\theta$
\be
{1\over 4}\omega_{\theta}^{\,~ij}\sigma_{ij} =-{1\over 2} H \gamma_{\theta}\ ,\quad H =- i\ .
\ee
We now determine the gauge connection that yields a nontrivial Killing spinor.
For a constant spinor, this relation implies that the Killing spinor equations (\ref{kseq1})(\ref{kseq2}) hold in the $\theta$ direction provided $A_\theta = 0$ modulo flat connection. By the same reasoning, the Killing spinor equation is solved provided $A_\phi =0$ modulo flat connection. Finally, with $H$ given as above, $A_\tau$ can be easily determined. Notice
\be
{1\over 4}\omega_\tau^{\,~ij}\sigma_{ij} +{1\over 2} H \gamma_\tau = {i\over 2}(1-q)\sigma_3\ ,
\ee
which gives the constant Killing spinor solution for (\ref{kseq1})\footnote{We take Killing spinors normalized. We further require that Killing solution should be invariant under $\tau\rightarrow \tau+2\pi$ due to the periodicity. Therefore we do not include those solutions depending on $q$.}
\be
\el{Ksp1}
\zeta = \left(
\begin{array}{c}
 0 \\
 1 \\
\end{array}
\right)\ ,
\ee with the gauge field $ A_\tau = {1 \over 2} (q-1)$. With this choice of $A_\tau$, the constant Killing spinor solution for (\ref{kseq2}) is
\be
\el{Ksp2}
\widetilde\zeta = \left(
\begin{array}{c}
 1 \\
 0 \\
\end{array}
\right)\ .
\ee
To summarize, we determined the supergravity backgrounds admitting two supercharges of opposite R-charge on $\mathbb{S}^3_q$:
\be\el{backgroundfield}
H(\mathbb{S}^3_q)= -i\ ,\qquad A(\mathbb{S}^3_q) = {1\over 2}\left(q-1\right)\rmd\tau\ , \qquad V(\mathbb{S}^3_q)=0\ .
\ee
This result was first obtained in~\cite{Nishioka:2013haa}. Here, we included our derivation to emphasize the strategy of finding Killing spinor solutions, which will be extended for more general $q$-branched spaces in subsequent sections.

So far we have been discussing the branched 3-sphere $\mathbb{S}^3_q$ \er{3sphere}, which has a conical singularity
at $\xt = 0$. As a common recipe \cite{Fursaev:1995ef} to handle the singularity, one may instead study a sequence of smooth resolved spaces $\mathbb{\widehat S}^3_q(\xe)$ ($\epsilon>0$ is small) and consider $\mathbb{S}^3_q$ as the $\xe \to 0$ limit of $\mathbb{\widehat S}^3_q(\xe)$. The metric of $\mathbb{\widehat S}^3_q(\xe)$ is given by
\begin{equation}
\rmd s^{2}= f_{\epsilon}\left(\theta\right)^2 \rmd\theta^{2}+q^{2}\ell^2\sin^{2}\theta \rmd\tau^{2}+\ell^2\cos^{2}\theta \rmd\phi^{2}\ ,\label{eq:resolved space}
\end{equation}
where $f_{\epsilon}\left(\theta\right)$ is a smooth function satisfying
\be
\el{resolvedf}
f_{\epsilon}\left(\theta\right)=\begin{cases}
q \ell \ , & \theta\rightarrow0 \ \\
\ell \ , & \epsilon<\theta\leq  \frac{\pi}{2} \ .
\end{cases}
\ee
%
One readily finds that the background fields permitting two supercharges with opposite
R-charge are \be\el{backgroundfield0}
H = -{i\over f_{\epsilon}(\theta)}\ ,~~A = {1\over 2}\left({q \ell\over f_{\epsilon}(\theta)}-1\right)\rmd\tau + {1\over 2}\left({\ell\over f_{\epsilon}(\theta)}-1\right)\rmd\phi \ , ~~V=0\ .
\ee
%
With the choice of vielbeins as in \er{eq:Resolved_space_vielbein}, the two Killing spinors are the same as \er{Ksp1} and \er{Ksp2} . As we shall discuss later, the partition function on the resolved space $Z[\mathbb{\widehat S}^3_q(\xe)]$ can be computed using the supersymmetry localization technique. In particular, the result does not depend on 
the specific form of the resolving function $f_\xe(\theta)$.

\subsubsection{$\widetilde{\mathbb{S}}^3_{p,q}$}
The study above can be extended to more general 3-spheres:  $(p,q)$-branched spaces, including branched ellipsoid, branched squashed sphere and the general branched sphere. Here, $p$ and $q$ are two conic deformation parameters of the two circles along $\phi$ and $\tau$
directions, respectively. For completeness, we include the results for each of these 3-spheres.

{\it Branched Ellipsoid}:~~~
The metric of $(p,q)$-branched 3-ellipsoid is given by
\be
\label{3ellipsoid}
\rmd s^2 = f(\theta)^2\rmd\theta^2 + p^2\ell^2 \cos^2\theta \rmd\phi^2 + q^2\tilde\ell^2\sin^2\theta \rmd\tau^2\ ,\quad f(\theta) = \sqrt{\ell^2\sin^2\theta+\tilde\ell^2\cos^2\theta}\ .
\ee
Following the procedure in section~\ref{S3q_kse} for $q$-branched round sphere, we find the Killing spinors remain the same:
\be
\zeta = \left(
\begin{array}{c}
 0 \\
 1 \\
\end{array}
\right)\ ,\quad \widetilde\zeta = \left(
\begin{array}{c}
 1 \\
 0 \\
\end{array}
\right)\ ,
\ee
with the supergravity background
\be\el{backgroundfield1}
H = -{i\over f(\theta)}\ ,~~A = {1\over 2}\left({q\tilde\ell\over f(\theta)}-1\right)\rmd\tau + {1\over 2}\left({p\ell\over f(\theta)}-1\right)\rmd\phi \ , ~~V=0\ .
\ee
In the limit $p \rightarrow 1$ and $\tilde{\ell} \rightarrow \ell$ (and under a replacement $f \to f_\epsilon$), 
the background \er{backgroundfield1} is reduced to (\ref{backgroundfield0}).

{\it Branched squashed sphere}:~~~
The metric for the smooth squashed 3-sphere is\footnote{Gamma matrices and the vielbein are listed in Appendix \ref{branchedsquashed_vielbein}. The same notation will be used in the one-loop computation for branched squashed sphere later.}
\be
\rmd s^2 =\ell^2\left({1\over v^2}\mu^1\mu^1 + \mu^2\mu^2 + \mu^3\mu^3\right)\ ,
\ee where $v$ is the squashing parameter. To make the $q$-branched space manifest, we go to ($\theta, \tau, \phi$) coordinates. We will set $\ell=1$ below. The metric can be written as
\be \rmd s^2 = \rmd\theta^2 + {1\over v^2}\left(\cos^4\theta \rmd\tau^2+\sin^4\theta\rmd\phi^2\right)+  \cos^2\theta\sin^2\theta (\rmd\tau^2+\rmd\phi^2) - {\sin^22\theta\over 2}\left(-{1\over v^2}+1\right)\rmd\phi \rmd\tau\ ,\ee
where the domains of $\theta, \tau, \phi$ are
\be
 \theta\in[0,\pi/2]\ ,\quad \tau\in [0,2\pi)\ , \quad \phi\in [0,2\pi)\ .
 \ee
 The $(p,q)$-branched squashed 3-sphere is obtained by replacing $(\rmd\phi, \rmd \tau)$ by $(p\rmd\phi, q\rmd\tau)$ in the metric, while keeping the domains of the coordinates intact
\be\label{branchedsquashed_metric} \rmd s^2 = \rmd\theta^2 + {1\over v^2} \left(\cos^4\theta q^2\rmd\tau^2+\sin^4\theta p^2\rmd\phi^2\right)+  \cos^2\theta\sin^2\theta (q^2\rmd\tau^2+p^2\rmd\phi^2) - {pq\sin^22\theta\over 2}\left(-{1\over v^2}+1\right)\rmd\phi \rmd\tau\ .\ee
For $p=q=1$, the space becomes  a squashed 3-sphere; for $v=1$, it becomes a $(p,q)$-branched, round 3-sphere.
Choosing the vielbein listed in Appendix~\ref{branchedsquashed_vielbein}, we found the same constant Killing spinor solutions as before
\be
\zeta = \left(
\begin{array}{c}
 0 \\
 1 \\
\end{array}
\right)\ ,\qquad \widetilde\zeta = \left(
\begin{array}{c}
 1 \\
 0 \\
\end{array}
\right)\ ,
\ee
with the following background fields
\be
\el{bgsquashed}
H = -{i \over v}\ ,~~A = \left(-\frac{q}{2 v^2}\left((v^2-1)\cos 2\theta-1\right) -{1\over 2} \right)\rmd\tau+\left(\frac{p}{2 v^2}\left((v^2-1)\cos 2\theta+1\right) -{1\over 2} \right)\rmd\phi\ , ~~V=0\ .
\ee

{\it A general 3-space with $U(1)\times U(1)$ isometry}:~~~
Having studied various branched spheres, we now move on to a general 3-space with $U(1)\times U(1)$ isometry. The space is characterized by three real parameters $p,q,v$ and one arbitrary function $f(\theta)$:
\be\label{generalmetric} \rmd s^2 = f(\theta)^2 \rmd\theta^2 + {1\over v^2} \left(\cos^4\theta q^2\rmd\tau^2+\sin^4\theta p^2\rmd\phi^2\right)
 +  \cos^2\theta\sin^2\theta (q^2\rmd\tau^2+p^2\rmd\phi^2) - {pq\sin^22\theta\over 2}\left(-{1\over v^2}+1\right)\rmd\phi \rmd\tau\ .\ee
Again, we find that this 3-space admits constant spinor solutions for the Killing spinor equations (\ref{kseq1})(\ref{kseq2}):
\be
\zeta = \left(
\begin{array}{c}
 0 \\
 1 \\
\end{array}
\right)\ ,\quad \widetilde\zeta = \left(
\begin{array}{c}
 1 \\
 0 \\
\end{array}
\right)\ ,
\ee with the background fields
\ba
H & = & -{i \over v f(\theta)}\ ,\quad ~~V\; =\; 0\ , \nn
A & = & \left(-\frac{q}{2 v^2f(\theta)}\left((v^2-1)\cos 2\theta-1\right) -{1\over 2} \right)\rmd\tau+\left(\frac{p}{2 v^2f(\theta)}\left((v^2-1)\cos 2\theta+1\right) -{1\over 2} \right)\rmd\phi\ .\nn
\ea
It can be shown that the metric \er{generalmetric} covers the round 3-sphere, 3-ellipsoid, squashed 3-sphere and their $(p,q)$-branched spaces, with different choices of parameters $p,q,v \in \mathbb{R}$ and functions $f(\theta)$. The general 3-space also covers more generally other singular and regular 3-spaces, in so far as the space preserves $U(1)\times U(1)$ isometry.

\subsection{Localization on branched spheres}
\label{weakcouplingpartition}
Consider an ${\cal N}=2$ supersymmetric field theory admitting Lagrangian formulation on a branched 3-space.
The partition function of the theory is invariant under the fermionic symmetries generated by the two supercharges $Q$ and $\widetilde Q$. 
These supersymmetries allow to evaluate the path integral by the localization technique: one adds a $Q$-exact localizing term $\{Q,V\}$ to the action. It follows from the supersymmetry algebra that the deformed partition function,
\begin{align}
Z(t) = \int \CD \phi \, e^{-S - t\{ Q, V\} } \ ,
\end{align}
is independent of $t$. The localization technique proceeds by choosing the bosonic part of $\{Q, V\}$ positive semi-definite and sending the deformation parameter $t \to \infty$ so that
\be
\{Q, V\}=0\
\ee
puts each independent positive semi-definite term to vanish. In the limit $t\to \infty$, the integral over critical points of $V$ (locus) can be evaluated exactly using the saddle-point approximation.
Once the field contents are specified, the explicit form of the deformation term $\{Q,V\}$ can be constructed from supersymmetry transformation rules, equivalently, the supersymmetric Lagrangian. Consider the ${\cal N}=2$ Chern-Simons-matter theory. The vector multiplet has components $(a_\mu, \xl, \bar \xl, \xs, D)$, transforming in the adjoint representation of the gauge group. The Yang-Mills term is $Q$-exact and can be used to localize the vector multiplet in the Coulomb branch
\be
\CL_{\text{YM}} = \text{Tr}\Bigg[\frac{1}{4}F_{\mu\nu}F^{\mu\nu}+\frac{1}{2}D_{\mu}\sigma D^{\mu}\sigma-i\bar{\lambda}\gamma^{\mu}D_{\mu}\lambda -\frac{1}{2}(D+\sigma H)^{2}-i\bar{\lambda}[\sigma,\lambda]+\frac{i}{2}H\bar{\lambda}\lambda\Bigg]\ ,
\ee
where
\begin{align}
\begin{aligned}
F_{\mu\nu} & :=  \partial_{\mu}a_{\nu}-\partial_{\nu}a_{\mu}-i[a_{\mu},a_{\nu}]\ , \\
D_{\mu}\sigma & :=  \partial_{\mu}\sigma-i[a_{\mu},\sigma]\ ,\\
D_{\mu}\lambda & :=  (\nabla_{\mu}+iA_{\mu})\lambda-i[a_{\mu},\lambda]\ .
\end{aligned}
\end{align}
The bosonic part of $\cal{L}_\text{YM}$
are positive semi-definite, so the path integral is localized to a matrix integral over the Coulomb branch
\begin{align}
\begin{aligned}a_\mu=0\ ,\qquad \sigma=\xs_0\ ,\qquad D=-H\sigma_0\ ,\end{aligned}
\end{align}
where $\xs_0$ is a Lie algebra valued constant matrix.
The integrand consists of saddle-point contribution and Gaussian fluctuations around the saddle-point. The latter is
a product of one-loop determinants of each dynamical fields. Only the Chern-Simons and Fayet-Iliopoulos (FI) terms contribute to the saddle-points
\ba
{\cal L}_{\text{CS}} &=& {k\over 4\pi} \Tr\left[ i \epsilon^{\mu\nu\rho}(a_\mu\partial_\nu a_\rho +{2i\over 3} a_\mu a_\nu a_\rho)-2D\sigma + 2i \bar\lambda\lambda \right]\ ,\label{CSL}\\
{\cal L}_{\text{FI}} &=& {\xi \over 2\pi}\Tr(D-\sigma H)\ .
\ea
For simplicity, we drop the FI term from now on --- their inclusion is straightforward and does not add any new features. The saddle-point contribution of the Chern-Simons term can be evaluated straightforwardly for different backgrounds.
The theory may contain chiral multiplet matter with components $(\phi, \psi, F)$ in arbitrary representations of the gauge group, but they are localized at the origin. Matter and gauge one-loop determinants will be computed explicitly for different backgrounds later.
The partition functions of $\CN =2$ Chern-Simons-matter theories on branched 3-spheres, obtained by the localization technique, take the form
\be
\label{Z}
Z[k, N; \mathfrak{g},\Delta;{\cal M}_3( b_1, b_2)] \, = \, \int [\rmd \sigma_0]
 \, e^{i k f(b_1, b_2) \mathrm{Tr}\, \sigma_0^2} \mbox{Det}_{\rm v} (\sigma_0, b_1, b_2; \alpha) \mbox{Det}_{\rm ch} (\sigma_0, b_1, b_2, \Delta; \rho)\ ,
\ee
where the three terms in the integrand are classical contribution, one-loop determinant of the vector multiplets, and one-loop determinant of the chiral multiplets. Possible nonperturbative terms are omitted since they are exponentially small in the large $N$ limit we are primarily interested in. The partition function depends on the coupling parameters $k, N$, on the Lie algebra $\mathfrak{g}$ of the gauge group, and on the geometric data $b_1, b_2$ of ${\cal M}_3$. So, 
$f(b_1, b_2)$ is a certain geometric function that depends on $b_1, b_2$, 
$\xa$ is the set of positive roots, $\rho$ is the weight space of the chiral multiplet, which is in a certain representation of the gauge group, and $\Delta$ is the conformal dimension of chiral supermultiplet fixed by the R-charge. Two geometric parameters $b_1$ and $b_2$ are determined by $p,q$ and the squashing parameters (\eg $v$). 
As we shall see later, $b_1$ and $b_2$ also specify the Reeb vectors on the branched spaces.\footnote{See section~\ref{Reeb_section} for the definition and discussions of Reeb vector.}

A particularly simplifying limit is the weak coupling limit $k \rightarrow 0$. In this case, the partition function is reduced to the chiral multiplet one-loop determinant at the origin of the Coulomb branch $\sigma_0 = 0$. Let us explain how this comes out. The classical contribution provides a Gaussian distribution to $\sigma_0$.  If we take $k\rightarrow\infty$ limit while keeping other geometric parameters fixed, we see that the classical contribution becomes a delta-function
\be
\lim_{k \rightarrow \infty} e^{i k f( b_1, b_2) \mathrm{Tr}\, \sigma_0^2} \quad \sim \quad \prod_{\rm Cartan} \delta (\sigma_0)\
\ee
up to normalization factors. We see that the partition function localized in the Coulomb branch becomes infinitely peaked at the origin $\sigma_0 = 0$ . At the origin, the one-loop determinant contribution of vector multiplet is reduced to unity. Intuitively, this follows from the fact that at the origin of the Coulomb branch vector multiplet is massless and supersymmetric cancellation between boson and fermion one-loop determinants ensures that the ratio is unity.  On the other hand, the one-loop determinant of the chiral multiplet depends on the geometric data through the R-charge dependence. This dependence continues to be present to the origin $\sigma_0 = 0$. Summarizing, at the weak coupling limit, we have
\be
Z[k, N; \mathfrak{g},\Delta;{\cal M}_3( b_1, b_2)] \quad {\longrightarrow} \quad \mbox{Det}_{\rm ch} (b_1, b_2; \Delta, \rho)
\quad \mbox{at} \quad {k \rightarrow \infty}.
\ee
In the following subsections, we shall check this assertion by explicit computations.

\subsubsection{Branched ellipsoid}\label{localization_branched_ellipsoid}
We compute the partition function of ${\cal N}=2$ Chern-Simons matter theory on branched ellipsoid background (\ref{3ellipsoid})(\ref{backgroundfield1}).
The saddle-point contribution from supersymmetric Chern-Simons term (\ref{CSL}) is
\be\el{saddle-squashed}
Z_{\text{saddle}}=e^{{i\pi k\over b_1b_2}\mathrm{Tr}\, \sigma_0^2}\ ,\quad b_1^{-1} = q\tilde\ell, \ b_2^{-1}=p\ell\ .
\ee

Now we compute the one-loop determinant of a chiral multiplet around 
the locus $\sigma=\sigma_0$.
The $Q$-exact term used to localize the matter fields is chosen as a total super-derivative~\cite{Hama:2011ea}
\be
\zeta{\widetilde \zeta}~{\cal L}_{\text{matter}} = \delta_{\zeta}\delta_{\widetilde \zeta} \left(\bar\psi\psi + 2 i\bar\phi\sigma\phi \right)\ .
\ee
This leads to the scalar kinetic operator
\be
\Delta_{\phi} = -D_\mu D^\mu - {2 i (\Delta-1)\over f(\theta)}v^\mu D_\mu + \sigma_0^2 + {2\Delta^2 - 3 \Delta\over 2 f(\theta)^2} + {\Delta R\over 4}\ ,
\ee
and the fermion kinetic operator
\be
\Delta_\psi = -i\gamma^\mu D_\mu - i\sigma_0 - {1\over 2f(\theta)} + {\Delta-1\over f(\theta)} \gamma^\mu v_\mu\ ,
\ee where $\Delta$ is R-charge of the scalar and $R$ is the positive Ricci scalar. The covariant derivatives are defined as
\ba
\el{covariantdmatter}
D_\mu \phi &=& (\nabla_\mu - i \Delta A_\mu )\phi\ ,\nn
D_\mu \psi &=& (\nabla_\mu - i (\Delta - 1)A_\mu )\psi\ .
\ea
The vector $v_\mu$ is defined as
\be
v_\mu = \zeta\gamma_\mu{\widetilde \zeta}\ ,
\ee where $\zeta = (0,1)^T,\ {\widetilde \zeta} = (1, 0)^T$ are two Killing spinors with R charge $+1$ and $-1$, respectively. The spinor product is defined as
\be
\zeta\lambda = \zeta^\alpha \varepsilon_{\alpha\beta} \lambda^\beta\ ,~~~\zeta\gamma^\mu \lambda = \zeta^\alpha(\varepsilon\gamma^\mu)_{\alpha\beta}\lambda^\beta\ ,
\ee where $\varepsilon_{\alpha\beta}$ is anti-symmetric $2\times 2$ matrix with non-vanishing components $\varepsilon_{12}=-\varepsilon_{21}=-1$.
Decomposing the scalar as \footnote{We use $\BZ$ to denote integers and $\BN$ for non-negative integers.}
\be
\phi(\theta,\tau,\phi) = \phi_0(\theta)e^{i m \tau + i n\phi}\ ,\qquad m,n\in\mathbb{Z}\ ,
\ee the equation of motion for the scalar is given by
\be
\Delta_{\phi} \phi = \lambda_s \phi\ .
\ee Decomposing the spinor as
\begin{align}
\psi(\theta,\tau,\phi)=e^{i(m\tau+n\phi)}\left(\begin{array}{c}
\psi_{1}(\theta)\\
e^{i(\tau+\phi)}\psi_{2}(\theta)
\end{array}\right)\ ,\qquad m,n\in\mathbb{Z}\ ,
\end{align} the equation of motion is given by
\be
\Delta_\psi \psi = \lambda_f \psi\ .
\ee
Spinor equations of motion can be decomposed to give a single second order ordinary differential equation for $\psi_1(\theta)$. If we map
\be
\psi_1(\theta) \sim \phi_0(\theta)\ ,
\ee their equations of motion are the same provided that the following matching condition is satisfied,
\be
 \lambda_s=\lambda_f(\lambda_f + 2 i \sigma_0) \ .
\ee
A scalar mode with eigenvalue $\lambda_s$ and a pair of fermion modes with eigenvalue $\lambda_f$ and $\lambda_f+2i\sigma_0$ will cancel with each other in the one-loop determinant as long as $\psi_1\neq 0$ and $\psi_2\neq 0$. The remaining contributions will come from those modes with only one of $\psi_1$ and $\psi_2$ vanishing. Denote the eigenvalue for $\psi_1\neq 0, \psi_2 =0$ as $\lambda_1$ and the eigenvalue for $\psi_1= 0, \psi_2\neq 0$ as $\lambda_2$. In the former case,
modes with $\lambda_1$ do not have pairing modes of $\lambda_1 + 2 i \sigma_0$.
In the latter case, there is no bosonic mode to cancel the fermionic modes with $\lambda_2$. $\lambda_1$ and $\lambda_2$ can be solved from spinor first order equations of motion. The remaining effective scalar mode gives the eigenvalue \footnote{Ranges of $m,n$ are determined by the normalizability condition. We emphasize that, resolving conditions at $\theta=0$ and $\pi/2$ are necessary to obtain the $(p,q)$ independent ranges of $m,n$.}
\be
\lambda_1+ 2 i\sigma_0 = {n\over p\ell} + {m \over q\tilde\ell} + {\Delta\over 2} \left( {1\over p\ell} + {1\over q\tilde\ell}\right) + i \sigma_0\ ,\quad(m,n\in \mathbb{N})
\ee
and the unmatched spinor eigenvalue is
\be
\lambda_2 = -{n\over p\ell} - {m \over q\tilde\ell} - {\Delta\over 2} \left( {1\over p\ell} + {1\over q\tilde\ell}\right)- i \sigma_0\ ,~~(m,n<0)
\ee
The one-loop determinant is given by
\be
\mbox{Det}_{\rm ch} ={\det \Delta_\psi\over \det \Delta_\phi} = \prod_{m,n\geq 0}  {{n\over p\ell} + {m \over q\tilde\ell} - {\Delta-2\over 2} \left( {1\over p\ell} + {1\over q\tilde\ell}\right)- i\sigma_0\over {n\over p\ell} + {m \over q\tilde\ell} + {\Delta\over 2} \left( {1\over p\ell} + {1\over q\tilde\ell}\right) + i\sigma_0}\ .\ee
Introducing familiar notations
\be
b = \sqrt{b_2\over b_1}:=b_0\sqrt{q\over p}\ ,~~b_0=\sqrt{ \tilde\ell\over\ell}\ , ~~ Q=b+1/b\ ,
\ee
we get
\be\label{Z_ellipsoid_branched}
\mbox{Det}_{\rm ch} (\sigma_0, b_1, b_2, \Delta; \rho) = s_b\left[{iQ(1-\Delta)\over 2} + {\rho(\sigma_0)\over \sqrt{b_1b_2}}\right]\ ,
\ee where $s_b(x)$ is double sine function.

Now we compute the one-loop determinant of the gauge fluctuations. For the fluctuations of Yang-Mills Lagrangian ${\cal L}_{\text{YM}}$ around the locus, one can impose the covariant gauge
\be
\nabla^\mu a_\mu = 0
\ee by adding the gauge fixing term
\be
{\cal L}_{\text{g.f.}} = \bar c\nabla_\mu D^\mu c + b\nabla^\mu a_\mu\ .
\ee
Decomposing the gauge potential
\be
a = B + d\chi\ , ~~\text{with}\quad\nabla^\mu B_\mu = 0\ ,
\ee the determinants from fluctuations $\chi$ and $\delta \sigma$ cancel with those from
the ghosts $c,\bar c$~\cite{Nishioka:2013haa}. The remaining gauge fixed Lagrangian for the fluctuations becomes
\be
{\cal L}_{\text{gauge}}= \Tr \left(B_\mu\Delta_B B^\mu-[B_\mu, \sigma_0]^2 - i \bar\lambda\gamma^\mu(\nabla_\mu + i A_\mu)\lambda + i \bar\lambda[\lambda, \sigma_0] +{\bar\lambda\lambda\over 2}\right)\ .
\ee
For all adjoint fields, one can decompose them with respect to the Cartan-Weyl basis
\be
[H_i, H_j] = 0,\quad [H_i, E_\alpha] = \alpha_i E_\alpha,\quad [E_\alpha, E_{-\alpha}]={2\alpha_i H_i\over |\alpha|^2}\ ,
\ee and the Lagrangian can be written as
\be
{\cal L}_{\text{gauge}} = \sum_i^r \biggr(B^i_\mu\Delta_B B^\mu_i + \bar\lambda_i\Delta_\lambda\lambda_i \biggr)+ \sum_\alpha \biggr(B_\mu^{-\alpha}(\Delta_B+\alpha(\sigma_0)^2)B^\mu_\alpha + \bar\lambda_{-\alpha}(\Delta_\lambda-i\alpha(\sigma_0))\lambda_\alpha\biggr)\ ,
\ee
where  $r$ is the rank of gauge group $G$ and $\sigma_0$ takes the value in the Cartan subalgebra. The kinetic operators $\Delta_B$ and $\Delta_\lambda$ are defined as ($\star$ for Hodge star operator)
\ba
\Delta_B &=& \star\,\rmd\star\rmd + \rmd\star\rmd\,\star\ ,\nn
\Delta_\lambda &=& -i\gamma^\mu(\nabla_\mu + i A_\mu) +{i \over 2}H\ .
\ea
Now we solve the eigenvalue problem for the vector Laplacian with a constraint~\cite{Klebanov:2011uf}
\be
\Delta_B B = \lambda_B^2 B\ ,~~\text{with}\quad \nabla^\mu B_\mu = 0\ ,
\ee which is equivalent to solve equations
\be
\rmd\star B=0\ ,\quad \star\,\rmd\,B=\lambda_B B\ .
\ee
We can decompose $B$ field in terms of the vielbein (\ref{eq:Resolved_space_vielbein})
\be
B = e^{i(m\tau+n\phi)}\left[e^{-i(\phi+\tau)}b_+(\theta)(e^1+ie^2)+e^{i(\phi+\tau)}b_-(\theta)(e^1-i e^2)+b_3(\theta)e^3\right]\ , \quad m,n\in \mathbb{Z}\ .
\ee
From $\star\,\rmd B=\lambda_B B$ we get
\be
b_{\pm}(\theta) = {1\over 2({m\over q\tilde\ell}+{n\over p\ell}\mp\lambda_B)}\left[\left({m\over q\tilde\ell}\cot\theta-{n\over p\ell}\tan\theta\right)b_3(\theta)\pm{b_3'(\theta)\over f(\theta)}\right]\ .
\ee Substituting it into $\rmd\star B=0$, we get a second order differential equation for $b_3(\theta)$
\be
\Delta_{b_3} b_3 = 0\ .
\ee
Decomposing spinor $\lambda$ as
\begin{align}
\lambda(\theta,\tau,\phi)=e^{i(m\tau+n\phi)}\left(\begin{array}{c}
\lambda_+(\theta)\\
e^{i(\tau+\phi)}\lambda_{-}(\theta)
\end{array}\right)\ ,\qquad m,n\in\mathbb{Z}\ ,
\end{align} the equation of motion
\be
(\Delta_\lambda-\lambda_f) \lambda =0\ ,
\ee  can be rewritten as a second order ordinary differential equation of $\lambda_+(\theta)$.
Note that equations of motion for $b_3(\theta)$ and $\lambda_+(\theta)$ coincide provided that the matching condition is satisfied
\be
(\lambda_B-\lambda_f)(\lambda_B+\lambda_f-{2\over f(\theta)}) = 0\ .
\ee
After cancellation of the matched eigenvalues, the remaining (effective) bosonic eigenvalue is
\be\label{gaugeDet_bosonic}
{2\over f(\theta)}-\lambda_1 = -{m\over q\tilde\ell}-{n\over p\ell}\ ,\quad (m,n\in \mathbb{N})
\ee
and the remaining fermionic eigenvalue is
\be
\lambda_2= -{m\over q\tilde\ell}-{n\over p\ell}\ .\quad (m,n< 0)
\ee
Note that the remaining eigenvalues are independent of $f(\theta)$ and we regularize the determinant by neglecting the zero mode in (\ref{gaugeDet_bosonic}). The one-loop determinant of the gauge fluctuations is
\ba
\label{gauge-Det}
\mbox{Det}_{\rm v} (\sigma_0, b_1, b_2; \alpha) &=&\left( {\prod_{m,n<0} -{m\over q\tilde\ell}-{n\over p\ell}\over \prod_{m,n>0} -{m\over q\tilde\ell}-{n\over p\ell}}\right)^r \prod_{\alpha>0}{\prod_{m,n<0} \left({m\over q\tilde\ell}+{n\over p\ell}\right)^2 + \alpha(\sigma_0)^2\over \prod_{m,n>0} \left({m\over q\tilde\ell}+{n\over p\ell}\right)^2+
\alpha(\sigma_0)^2}\nn
&\sim& \prod_{\alpha>0}\left[ {1\over \alpha(\sigma_0)^2}\times 4\sinh \frac{\pi\alpha(\sigma_0)}{b_1}\sinh\frac{\pi\alpha(\sigma_0)}{b_2}\right]\ .
\ea
In the last step, we dropped an overall constant, which is irrelevant for the discussion. Combining \er{saddle-squashed}, \er{Z_ellipsoid_branched} and \er{gauge-Det} the total partition function is given by
\be
\label{Z}
Z[k, N; \mathfrak{g},\Delta;b_1, b_2] \, = \, \int \prod_{i=1}^r\rmd(\sigma_0)_i \, 
e^{{i\pi k\over b_1b_2} \mathrm{Tr}\, \sigma_0^2} 
\prod_{\alpha>0} 4 \sinh \frac{\pi\alpha(\sigma_0)}{b_1}\sinh\frac{\pi\alpha(\sigma_0)}{b_2}\prod_{\rho}s_b\left(\frac{i Q}{2}(1-\xD)+\frac{\rho(\sigma_0)}{\sqrt{b_1b_2}}\right)\ ,
\ee
where $r$ is the rank of the gauge group and $(\sigma_0)_i$ denote the Cartan parts of $\sigma_0$. Note that ${1\over \alpha(\sigma_0)^2}$ in the gauge determinant will cancel the Vandermonde determinant in the measure, therefore we get the final result (\ref{Z}) shown above. 
The partition function on the $(p,q)$-branched ellipsoid is the same as that on the smooth ellipsoid with redefined squashing $\ell\to p\ell, \tilde\ell\to q\tilde\ell$. Particularly, in the round sphere limit $\tilde\ell=\ell$, it will be the same as that on the smooth ellipsoid with $b=\sqrt{q\over p}$~\cite{Nishioka:2013haa}.
Note that the full partition function is independent of the specific form of $f(\theta)$, which shows that the result is valid for arbitrary $f(\theta)$.

\subsubsection{Branched squashed sphere}\label{localization_branched_squashed}
We compute the partition function of ${\cal N}=2$ Chern-Simons matter theory on the branched squashed sphere background (\ref{branchedsquashed_metric})(\ref{bgsquashed}).
The saddle-point contribution is
\be
Z_{\text{saddle}}=e^{{i\pi k\over b_1b_2}\mathrm{Tr}\, \sigma_0^2}\ ,\quad b_1 = {v\over q}\ , \ b_2={v\over p}\ .
\ee
Now we compute the one-loop determinant of a chiral multiplet. 
The $Q$-exact term used to localize the matter fields is chosen to be a total super-derivative~\cite{Hama:2011ea}
\be
\zeta{\widetilde \zeta}~{\cal L}_{\text{matter}} = \delta_{\zeta}\delta_{\widetilde \zeta} \left(\bar\psi\psi + 2 i\bar\phi\sigma\phi +{2(1-\Delta)\over v}\bar\phi\phi \right)\ .
\ee
This leads to the operators
\ba
\Delta_{\phi} &=& -D_\mu D^\mu + \sigma_0^2 + {\Delta(1 - 2 \Delta)\over 2 v^2} + {2i(1-\Delta)\sigma_0\over v}+ {\Delta R\over 4}\ ,\\
\Delta_\psi &=& -i\gamma^\mu D_\mu - i\sigma_0 - {2\Delta-1\over 2v}\ ,
\ea where $\Delta$ is the R-charge of the scalar component field. Decomposing the scalar field as
\be
\phi(\theta,\tau,\phi) = \phi_0(\theta)e^{i m \tau + i n\phi}\ ,\qquad m,n\in\mathbb{Z}\
\ee the equation of motion is given by
\be
\Delta_{\phi} \phi = \lambda_s \phi\ .
\ee Decomposing the spinor field as
\begin{align}
\psi(\theta,\tau,\phi)=e^{i(m\tau+n\phi)}\left(\begin{array}{c}
e^{i(-\tau-\phi)}\psi_{1}(\theta)\\
\psi_{2}(\theta)
\end{array}\right)\ ,\qquad m,n\in\mathbb{Z}\ ,
\end{align} the equation of motion is given by
\be
\Delta_\psi \psi = \lambda_f \psi\ .
\ee
As in the case of branched ellipsoid, the equations of motion for $\psi_2(\theta)$ and $\phi_0(\theta)$ are the same when
\be
\text{$\lambda_s $}= \left(\lambda_f +{1\over v}\right)\left(\lambda_f+{2\Delta -1\over v} + 2 i \sigma_0\right)\ .\ee
According to the analysis in the previous subsection, after cancellation, the remaining eigenvalues are (bosonic)
\be
(1+ v \lambda_1)/v = {m v\over q}+{n v\over p}-{\Delta  v\over 2}\left({1\over p}+{1\over q}\right) - i \sigma_0  \ ,~~(-m,-n\in \mathbb{N})
\ee
and (fermionic)
\be
(2 \Delta +v (\lambda_2+2 i \sigma_0)-1)/v=-{m v\over q}-{n v\over p}+{\Delta  v\over 2}\left({1\over p}+{1\over q}\right) + i \sigma_0\ .~~(-m,-n<0)
\ee
Putting a minus sign in front of both the numerator and denominator, we obtain the final one-loop determinant
\ba
\mbox{Det}_{\rm ch} = {\det \Delta_\psi\over \det \Delta_\phi} = \prod_{m,n\geq 0}  {{nv\over p} + {mv \over q} - {\Delta-2\over 2} \left( {v\over p} + {v\over q}\right) -i \sigma_0 \over {nv\over p} + {mv \over q} + {\Delta\over 2} \left( {v\over p} + {v\over q}\right) + i \sigma_0}\\
=\prod_{m,n\geq 0}  {{n\over p} + {m \over q} - {\Delta-2\over 2} \left( {1\over p} + {1\over q}\right) -{i\sigma_0\over v} \over {n\over p} + {m \over q} + {\Delta\over 2} \left( {1\over p} + {1\over q}\right) + {i \sigma_0\over v}}\ .\ea
Introducing the notations
\be
b = \sqrt{b_2\over b_1}\ ,~~ Q=b+1/b\ , \quad b_1 = {v\over q}\ , \ b_2={v\over p}\ .
\ee
we get
\be\label{Z_squashed_branched}
\mbox{Det}_{\rm ch} (\sigma_0, b_1, b_2, \Delta; \rho) = s_b\left[{iQ(1-\Delta)\over 2}+{\rho(\sigma_0) \over \sqrt{b_1b_2}}\right]\ .
\ee
The one-loop determinant of the gauge fluctuations can be computed as before. 
One needs to solve the vector eigenvalue problem
\be
\rmd\star B=0\ ,\quad \star\,\rmd\,B=\lambda_B B\ ,
\ee and the spinor eigenvalue problem
\be
\left[-i\gamma^\mu (\nabla_\mu + i A_\mu) + {1\over 2v}-\lambda_f\right]\lambda = 0\ .
\ee
$B$ field can be decomposed in terms of the vielbein (\ref{squashed_vielbein})
\be
B = e^{i(m\tau+n\phi)}\left[e^{-i(\phi+\tau)}b_+(\theta)(e^2+ie^3)+e^{i(\phi+\tau)}b_-(\theta)(e^2-i e^3)+b_0(\theta)e^1\right]\ , \quad m,n\in \mathbb{Z}\ .
\ee
From $\star\,\rmd B=\lambda_B B$ we get
\be
b_{\pm}(\theta) = {1\over 2({mv\over q}+{nv\over p}\pm\lambda_B)}\left[\left({m\over q}\cot\theta-{n\over p}\tan\theta\right)b_0(\theta)\pm b_0'(\theta)\right]\ .
\ee
Substituting it into $\rmd\star B=0$, we get a second-order differential equation for $b_0(\theta)$
\be
\Delta_{b_0} b_0 = 0\ .
\ee
We decompose spinor $\lambda$ as
\begin{align}
\lambda(\theta,\tau,\phi)=e^{i(m\tau+n\phi)}\left(\begin{array}{c}
e^{i(-\tau-\phi)}\lambda_{+}(\theta)\\
\lambda_{-}(\theta)
\end{array}\right)\ ,\qquad m,n\in\mathbb{Z}\ .
\end{align}
and obtain a second order ordinary differential equation for $\lambda_-(\theta)$. Again equations of motion for $b_0(\theta)$ and $\lambda_-(\theta)$ coincide if
\be
(\lambda_B-\lambda_f+{1\over v})(\lambda_B+\lambda_f+{1\over v}) = 0\ .
\ee
We then have the uncanceled eigenvalues (bosonic)
\be
\lambda_1-{1\over v} = -\frac{m v}{q}-\frac{n v}{p}\ ,\quad (-m,-n\in\mathbb{N})
\ee
and (fermionic)
\be
-(\lambda_2+{1\over v}) = -\frac{m v}{q}-\frac{n v}{p}\ .\quad (-m,-n<0)
\ee
The one-loop determinant for the gauge fluctuations is
\ba
\mbox{Det}_{\rm v} (\sigma_0, b_1, b_2; \alpha) &=&\left( {\prod_{m,n<0} {mv\over q}+{nv\over p}\over \prod_{m,n>0} {mv\over q}+{nv\over p}}\right)^r \prod_{\alpha>0}{\prod_{m,n<0} \left({mv\over q}+{nv\over p}\right)^2 + \alpha(\sigma_0)^2\over \prod_{m,n>0} \left({mv\over q}+{nv\over p}\right)^2+
\alpha(\sigma_0)^2}\nn
&\sim&\prod_{\alpha>0}\left[ {1\over \alpha(\sigma_0)^2}\times 4\sinh \frac{\pi\alpha(\sigma_0)}{b_1}\sinh\frac{\pi\alpha(\sigma_0)}{b_2}\right]\ .
\ea
Note that $b_1,b_2$ now have different physical meanings from those of the branched ellipsoid
\er{saddle-squashed}.
However, up to an overall constant the partition function on the $(p,q)$-branched squashed sphere is the same as that on the $(p,q)$-branched round sphere. Notice that $v$-type squashing does not affect the partition function even for branched sphere \footnote{It was shown in \cite{Hama:2011ea} that the partition function on squashed sphere (\er{branchedsquashed_metric} with $p=q=1$) remains the same as that on round sphere.}.

\subsection{Reeb vector and parameter dependence of $Z_{{\cal M}_3}$}
\label{Reeb_section}
All of the backgrounds we discussed in subsection~\ref{subsec:Killing} admit at least two supercharges with opposite R-charges and also have at least $U(1) \times U(1)$ isometry. In case the base space is smooth, it has a toric contact structure. The associated Killing Reeb vector field $K$, which can always be constructed from bilinear of Killing spinors
\be
K = \zeta \gamma^\mu {\widetilde \zeta} \partial_\mu\ ,
\ee
can be expressed as the linear combination of the two $U(1)$ Killing vectors,
\be
\el{Reebvector} K= b_1\partial_\tau + b_2\partial_\phi\ .
\ee
Recently it was shown ~\cite{Alday:2013lba} that the partition function $Z_{{\cal M}_3}$ of $\CN =2$ Chern-Simons-matter theories on a 3-manifold with $U(1)\times U(1)$ isometry (and the topology of $\mathbb{S}^3$) can be computed using the supersymmetry localization technique. The result is exactly the same as (\ref{Z}) \footnote{Partition functions of the same form were previously obtained for ellipsoid~\cite{Hama:2011ea} and squashed sphere \cite{Imamura:2011wg}. For recent developments on localization on three sphere or deformed three spheres, see~\cite{Jafferis:2010un,Hama:2010av,Imamura:2011uw,Nian:2013qwa,Tanaka:2013dca}.} with $b_1, b_2$ being the parameters of the Reeb vector \er{Reebvector}. 
We show that the same form of (\ref{Z}) holds even for singular spaces such as the branched spheres, where metrics are singular but Reeb vectors are still regular (specified by $b_1$ and $b_2$).
It can be seen from (\ref{Z}) that $Z_{{\cal M}_3}$ only depends on a single parameter
\be\el{bpara} b = \sqrt{b_2\over b_1}\ .\ee
Because a rescaling of both $b_1$ and $b_2$ by a constant only contributes an overall constant to the matrix integral by a redefinition of the integration variable $\xs_0$.

\subsubsection{Parameter Dependence of $Z_{{\cal M}_3}$}
The assertion that $Z_{{\cal M}_3}$ depends only on the ratio $b$ can also be understood without explicit computation. Here
we first recapitulate the relevant results from ~\cite{Closset:2013vra}
and then explain why all the partition functions 
(some of which have singular spaces as their limits) discussed in subsection~\ref{subsec:Killing} have the same form. For readers who are not interested in the details, the short answer is the following. First, we can consider all these examples as deformations of round sphere. All the deformations in geometry (including metric, almost contact structure etc.) other than $\xT$ (a quantity built from deformations
in the almost contact structure, see \er{Theta} below) only give $Q$-exact terms in the Lagrangian
and therefore do not contribute to the partition function. Finally, $\xT$ \er{xTeta} is entirely determined by the Reeb vector $\xi^\mu$.

Supersymmetric field theories on 3-manifolds have been studied in great details in \cite{Closset:2012ru}. In order to have a single supercharge, the manifold $\CM$ must admit an integrable almost contact structure. An almost contact structure is defined by a vector~$\xi^\mu$, a one-form~$\eta_\mu$, and an endomorphism~${\Phi^\mu}_\nu$, which satisfy,
\be\el{eq:contactstructure} {\Phi^\mu}_\rho {\Phi^\rho}_\nu = - {\delta^\mu}_\nu + \xi^\mu \eta_\nu~, \qquad \eta_\mu \xi^\mu =1~.\ee
Note that $\eta_\mu$ and $\xi^\mu$ are left and right kernels of ${\Phi^\mu}_\nu$, respectively. The almost contact structure can be understood as an odd dimensional analogue of complex structure. In the subspace
of tangent bundle orthogonal to $\eta_\mu$, ${\Phi^\mu}_\nu$ serves as an almost complex structure as $\Phi^2 = -1$ in this subspace.

More explicitly, with the integrable almost contact structure, we can define a projection operator
\be {\Pi^\mu}_\nu = \frac 1 2 \left({\delta^\mu}_\nu - i {\Phi^\mu}_\nu - \xi^\mu \eta_\nu\right)~, \qquad {\Pi^\mu}_\nu {\Pi^\nu}_\rho = {\Pi^\mu}_\rho\ ,\ee
and use it to separate the complexified tangent and cotangent bundles into holomorphic and anti-holomorphic subspaces. Holomorphic vectors~$X \in T^{1, 0} \CM$ and holomorphic one-forms~$\omega^{1,0} \in \Lambda^{1,0}$ are defined by
\be {\Pi^\mu}_\nu X^\nu = X^\mu~, \qquad \omega^{1,0}_\mu {\Pi^\mu}_\nu = \omega^{1,0}_\nu\ .\ee
Anti-holomorphic one-forms~$\omega^{0,1} \in \Lambda^{0,1}$ are defined by
$\omega^{0,1}_\mu {\Pi^\mu}_\nu = 0$.

The space of complex $k$-forms  $\Lambda^k$ can be decomposed into $(p,q)$-forms defined
by $\Lambda^{p,q} = \wedge^p \Lambda^{1,0} \otimes \wedge^q \Lambda^{0,1}$. Similar to the Dolbeault
operators $\bar \p$ on a complex manifold, we can define a nilpotent operator $\widetilde \p$ by projecting the exterior derivative of a $(p,q)$-form $\omega^{p,q}$ to $\Lambda^{p,q+1}$ (normally $\rmd \omega^{p,q} \in \Lambda^{p+1,q} \oplus \Lambda^{p,q+1}$):
\be\widetilde \p : \Lambda^{p,q} \rightarrow \Lambda^{p,q+1}~, \qquad \widetilde \p \omega^{p,q} = \rmd \omega^{p,q} \big|_{\Lambda^{p,q+1}}~.\ee
The relation $\widetilde \p^2 =0$ follows from $\rmd^2 =0$ and we have the $\widetilde \p$-cohomology,
\be H^{p,q}(\CM) = \frac { \{ \omega^{p,q} \in \Lambda^{p,q} | \widetilde \p \omega^{p,q} = 0 \} } {\widetilde \p \Lambda^{p, q-1}}~.\ee

The almost contact structure becomes integrable if $\xi^\mu$, $\eta_\mu$, ${\Phi^\mu}_\nu$ satisfy
\be {\Phi^\mu}_\nu {\left(\CL_\xi \Phi\right)^\nu}_\rho = 0~,\ee
where $\CL_\xi$ is the Lie derivative along $\xi^\mu$. In this case, we can have adapted charts $(\tau, z, \bar z)$ covering
the manifold so that in each patch $\xi^\mu$, $\eta_\mu$, ${\Phi^\mu}_\nu$ are in the form of,
\be \el{adaptedcoord}\xi = \p_\tau~, \qquad \eta = \rmd\tau + h \rmd z + \bar h \rmd {\bar z}~, \qquad {\Phi^\mu}_\nu = \bay{ccc}0 & - i h  & i \bar h \cr 0 & i & 0 \cr 0 & 0 & -i \eay~, \ee
where~$h(\tau, z, \bar z)$ is a complex function.
Coordinates in overlapping patches are related to each other by $\tau' = \tau + t(z,\bar z), z' = f(z)$,
where $f(z)$ is a holomorphic function and $t(z,\bar z)$ is real. Note that we have
$\Phi^z{}_{z} = -\Phi^{\bar z}{}_{\bar z} = i, \Phi^z{}_{\bar z} = \Phi^{\bar z}{}_{ z} = 0$,
which is like the complex structure in its canonical form.
Holomorphic vectors and one-forms read in adapted coordinates as
\be X = X^z (\p_z - h \p_\tau) ~, \qquad \omega^{1,0} = \omega^{1,0}_z \rmd z ~,\ee
while anti-holomorphic one-forms read as
\be\omega^{0,1} =\omega^{0,1}_\tau \left(\rmd\tau + h \rmd z\right) +\omega^{0,1}_{\bar z}\rmd \bar z~.\ee

We can always find a compatible metric for a certain almost contact structure
$\xi^\mu$, $\eta_\mu$, ${\Phi^\mu}_\nu$,
\be g_{\mu\nu} {\Phi^\mu}_\alpha {\Phi^\nu}_\beta  = g_{\alpha\beta} - \eta_\alpha \eta_\beta~.\ee
Given this metric we can express $\eta_\mu$ and ${\Phi^\mu}_\nu$ in terms
of $\xi^\mu$,
\be\eta_\mu = g_{\mu\nu} \xi^\nu,\quad  {\Phi^\mu}_\nu = - {\xe^\mu}_{\nu\rho} \xi^\rho\ .\ee
In adapted coordinates, the compatible metric is in the form of
\be \rmd s^2 = \eta^2 + c\left(\tau, z, \bar z\right)^2 \rmd z \rmd \bar z = \left(\rmd \tau + h\left(\tau, z, \bar z\right) \rmd z + \bar h\left(\tau, z, \bar z\right)\rmd \bar z\right)^2 + c\left(\tau, z, \bar z\right)^2 \rmd z \rmd\bar z~.\ee
In the following discussion compatible metric is assumed.

Now we can consider deformations of background fields and study the dependence of the partition function. At linearized level, the coupling between the background fields and the R-multiplet takes the form (only bosonic sector displayed)
\be \Delta \CL=-\frac 1 2 \Delta g^{\mu\nu} T_{\mu\nu} + A^{\mu} j_\mu^{( R)}+C^\mu j_\mu^{(Z)}+HJ^{(Z)}~,\ee
where $T_{\mu\nu}$ is the stress tensor and $j_\mu^{( R)}, j_\mu^{(Z)}, J^{(Z)}$
are operator components in the R-multiplet. $C^\mu$ is related to $V_\mu$ by
eq.\er{strengths_V_H}. Note that the background fields $A_\mu, V_\mu$ and $H$ are determined by the almost contact structure and the metric.
This is essentially how to show that the existence of almost contact structure allows supersymmetric background.
Not all the components of $\xD \xi^\mu$, $\xD \eta_\mu$, $\xD {\Phi^\mu}_\nu$
and $\xD g_{\mu\nu}$ are independent. The deformed contact structure needs to satisfy (\er{eq:contactstructure}). There are also constraints from metric being compatible and the integrability condition of the almost contact structure. It can be shown that all the other components are fixed by $\Delta \xi^\mu$, $\Delta \eta_z$, $\Delta \eta_{\bar z}$, $\Delta {\Phi^\tau}_\tau$, $\Delta {\Phi^z}_{\bar z}$, $\Delta {\Phi^{\bar z}}_z$ and $\xD g_{z \bar z}$ and there are additional constraints on $\xD {\Phi^\tau}_\tau, \Delta {\Phi^z}_{\bar z}, \Delta \xi^z$ and $\Delta \xi^{\bar z}$ (see below) from the integrability condition.

Moreover, most of the coefficients of the geometry deformations in $\Delta \CL$, which are linear combinations of the components in the R-multiplet, are $Q$-exact operators and therefore do not affect the partition function. Only $\Delta {\Phi^z}_{\bar z}$ and $\Delta \xi^z$ provide nontrivial deformations. Yet they are not independent as integrability condition imposes the constraint $\widetilde \p \xT^z = 0$, where $\Theta^z$ is built from $\Delta {\Phi^z}_{\bar z}$ and $\Delta \xi^z$ and it is the $z$-component of a $(0,1)$-form with coefficients in the holomorphic tangent bundle~$T^{1,0}\CM$,
\be \el{Theta}\Theta^z :=  - 2 i \Delta \xi^z \left(\rmd\tau + h \rmd z\right) + \left(\Delta {\Phi^z}_{\bar z} - i \bar h \Delta \xi^z\right) \rmd{\bar z}~. \ee
A trivial deformation of the almost contact structure due to an infinitesimal diffeomorphism (generated by $\xe^\mu$) corresponds to an exact $\Theta^z = 2 i \widetilde \p \epsilon^z$
\be \Delta \xi^z = - \p_\tau \xe^z~, \qquad \Delta {\Phi^z}_{\bar z} = 2 i \p_{\bar z} \xe^z - i \bar h \p_\tau \xe^z\ ,
\ee
and therefore only the cohomology class of $\Theta$ can affect the partition function. Summarizing, the partition function only depends on the cohomology class of~$\Theta$ in~$H^{0,1}\left(\CM, T^{1,0}\CM \right)$.

Let us now focus on the cases we are interested in, namely manifolds with
the topology of an $\mathbb{S}^3$. It can be shown that every element in $H^{0,1}\left(\mathbb{S}^3, T^{1,0}\mathbb{S}^3 \right)$ can be expressed (up to an exact form) in the form of
\be \el{xTeta}\xT = \xg X \otimes \eta\ ,\ee
where $\xg$ is a complex-valued deformation parameter and $X$ is a holomorphic vector fields $X = X^z(\p_z - h \p_\tau)$ satisfying $\widetilde \p X = 0$. As we shall see, resolved sphere, ellipsoid and squashed sphere \footnote{To have nontrivial deformation on squashed sphere, one needs background fields different from those given in eq.\er{bgsquashed}. See \cite{Imamura:2011wg} \cite{Closset:2013vra} for more details.} all give $X = z(\p_z - h \p_\tau)$. In fact, this is the case for the generic $U(1) \times U(1)$ manifold discussed above. The only difference in their $\xT$ 
is the complex parameter $\xg$, which is what the corresponding partition functions should depend on. Obviously there is an one to one correspondence between $\xg$ and $b$.

It is not difficult to compute the vector $X$ explicitly. In these cases, $\xi$
is the Reeb vector $K$. For simplicity, we consider $K = {1\over q}\partial_\tau + \partial_\phi$. In the adapted coordinates \footnote{To avoid confusion, here we use $\psi$ instead of $\tau$ and $\xi = \p_\psi$.},
\be \psi = \phi + \tau,\quad z = \tan \xt e^{i(\phi-\tau)}\ , \ee
the metric of a round sphere takes the following form,
\be \rmd s^2=\eta^2+{4\rmd z \rmd\bar z\over (1+ |z|^2)^2}\ , \qquad \eta=\rmd\psi+{i\over 2} {(\bar z \rmd z - z \rmd \bar z)(1 - |z|^2) \over |z|^2(1+ |z|^2) }\ .
\ee
The Reeb vector $K$ corresponds to a deformation in $\xi$ from $\xi = \partial_\tau + \partial_\phi$ on a round sphere (ellipses denote other components of $\xD \xi$),
\[\xD \xi = \lb \frac 1 q - 1\rb \p_\tau = \lb \frac 1 q - 1\rb\lb -i z \p_z
+ \dots\rb\ , \]
which implies
\be \xD \xi^z = -i \lb \frac 1 q - 1\rb z\ .\ee
From the definition of $\xT$ \er{Theta} and the fact \er{xTeta} that it is proportional
to the 1-form $\eta$ \er{adaptedcoord} \footnote{We can check that $\xD \Phi^z{}_{\bar z}
= -i \bar h \xD \xi^z + \frac i 2 c^2
\xD g^{zz} = -i \bar h \xD \xi^z - i \p_{\bar z} z'$, where the second term is a total derivative and corresponds to an exact form. In the second equality, we use the
fact that in the adapted coordinates $(\tau',z',\bar
z')$ for the deformed sphere, $g^{zz} = 0$ and $g^{z\bar z} = 2/c^2$ (the second equation
is true up to zeroth order in the deformation parameters).}, we get
\be X^z = z\ ,\quad\xg = 2\left(1-\frac
1 q \right)\ .
\ee

\subsubsection{Generalization to branched spaces}
The conclusion that partition function only depends on the Reeb vector holds even for branched spheres. This is because insertion of a conical singularity to an otherwise smooth manifold with almost contact structure does not break $U(1)\times U(1)$ isometry and because addition of appropriate background gauge field compensates the curvature singularity for the Killing spinors. One way to reach to this conclusion is to use the resolution argument. The argument goes as follows. Since the resolved  space is smooth, the partition function $Z_\epsilon:=Z[\hat{\mathbb{S}}^3_q(\xe)]$ will only depend on the Reeb vector. A key observation is that, Reeb vector is regular even for the limit $\epsilon\to 0$, where the space becomes singular. We further notice that, Reeb vector does not depend on the resolving factor $f_\epsilon(\theta)$ and therefore does not depend on the small parameter $\epsilon$. This chain of dependence relations gives
\be
\partial_\epsilon Z_\epsilon = 0\ .
\ee
With the assumption that $Z_\epsilon$ is a smooth function of $\epsilon$, we conclude that the partition function on $\mathbb{S}_q^3$
\be
Z_q=Z_{\epsilon\to 0} = Z_{\epsilon>0}\ .
\ee
The fact that partition function only depends on Reeb vector on both smooth spheres and branched spheres actually tells us that partition functions of supersymmetric field theories on all the backgrounds we discussed in~\ref{subsec:Killing} share the same form of (\ref{Z}), with $b$ defined from the Reeb vectors.

We can also check this conclusion by direct computations. As we discussed in section~\ref{weakcouplingpartition}, the supersymmetry localization technique is also applicable to the branched spaces. The partition functions take the form of (\ref{Z}). We have already studied Killing spinor equations in various branched spheres explicitly, so we simply list the Reeb vectors in which the overall size $\ell$ is also restored:

{\it Round sphere}
\be
K\ell= \partial_\tau + \partial_\phi\ ,
\ee

{\it Ellipsoid}
\be\label{Reeb_ellipsoid}
K\ell= {\ell\over \tilde\ell}\partial_\tau + \partial_\phi\ ,
\ee

{\it Squashed sphere}
\be\label{Reeb_squashed}
K\ell = v\partial_\tau + v\partial_\phi\ ,
\ee

{\it Branched round sphere}
\be
K\ell = {1\over q}\partial_\tau + {1\over p}\partial_\phi\ ,
\ee

{\it Branched ellipsoid}
\be
K\ell= {\ell\over q\tilde\ell}\partial_\tau + {1\over p}\partial_\phi\ ,
\ee

{\it Branched squashed sphere}
\be
K\ell = {v\over q}\partial_\tau + {v\over p}\partial_\phi\ ,
\ee
{\it The general metric with $U(1)\times U(1)$}
\be
K\ell = {v\over q}\partial_\tau + {v\over p}\partial_\phi\ .
\ee
Based on the Reeb vector results, we have the following observations
\begin{itemize}
\item
Squashed sphere shares the same partition function with round sphere. This was first pointed out in~\cite{Hama:2011ea} by explicit  computation using localization.
\item
Branched round sphere shares the same partition function with the ellipsoid by the identification
\be
{q\over  p}= {\tilde\ell\over \ell}\ .
\ee This was also observed in~\cite{Nishioka:2013haa} explicitly by localization computation.
\item Branched ellipsoid shares the same partition function with ellipsoid by redefining the squashing
\be
\ell \rightarrow p\ell\ ,\quad \tilde\ell \rightarrow q\tilde\ell\ .
\ee This was also observed in \ref{localization_branched_ellipsoid} explicitly by localization computation.
\item Branched squashed sphere shares the same partition function with branched round sphere, therefore with ellipsoid as well. This was also observed in \ref{localization_branched_squashed} explicitly by localization computation.
\item The general three-dimensional space with $U(1)\times U(1)$ symmetry shares the same partition function as branched round sphere.
\end{itemize}

Notice that Reeb vector does not depend on $f(\theta)$ in any of these cases.
For $\mathbb{S}^3_q$, there could be many different ways to resolve the singularity without changing the Reeb vector, which would lead us to many different resolved 3-spheres described by different $f(\theta)$. But all of them share the same partition function. 

\subsection{Partition function in the large $N$ limit}
In the large $N$ limit, the exact result of the partition function simplifies further.
Because $\mathbb{S}^3_q$ and $\widetilde{\mathbb{S}}^3_b$ share the same Reeb vector, we can identify the supersymmetric partition function on $\mathbb{S}^3_q$ with the one on $\widetilde{\mathbb{S}}^3_b$ :
\be
Z_q = Z[\widetilde{\mathbb{S}}^3_b] \Big\vert_{b = \sqrt{q}}\ .
\ee
The latter can be solved in the large $N$ limit (while holding other parameters fixed) as in~\cite{Imamura:2011wg, Martelli:2011fu, Martelli:2011fw, Martelli:2012sz, Martelli:2013aqa}
\be
\log Z[\widetilde{\mathbb{S}}^3_b] =  {1\over 4}\left(b+{1\over b}\right)^2\log Z_{b=1}\ .
\ee
Therefore, we have the partition function of qSCFT$_3$
\be\label{qpartition}
\log Z_q = {(q+1)^2\over 4q}\log Z_1\ .
\ee
By the definition
\be\el{SReyni} S_q = {q \log Z_1-\log Z_q \over q-1}\ ,
\ee we get the R\'enyi entropy
\be \el{super-Reyni-ft}
S_q = {3q+1\over 4q}S_1\ , \quad S_1=\log Z_1=-F_1\ ,
\ee
where $S_1$ is the entanglement entropy (EE), which is defined as the $q\to 1$ limit of the R\'enyi entropy. \footnote{$S_1=-F_1$ can be considered due to the fact $\partial_q (\log Z_q)|_{q\to 1}=0$, from the other equivalent definition of R\'enyi entropy $S_q=-\partial_q\log Z_q +\log Z_1$. This relation between entanglement entropy and free energy on $\mathbb{S}^3$ for general CFTs in any coupling was proved in \cite{Casini:2011kv}.}
The result is remarkably simple, factorizing out the branching parameter dependence.

Factorization of the R\'enyi entropy was first observed in~\cite{Nishioka:2013haa} for branched round sphere. As we discussed in the last subsection, because the Reeb vector remains the same, this formula holds even for branched ellipsoid, branched squashed sphere, and general spaces with $U(1)\times U(1)$ isometry (\ref{generalmetric}) with proper definitions of the effective parameter $b$ (or$\sqrt q$).

\subsection{From CFT on $\mathbb{S}^3_q$ to CFT on $\mathbb{S}^1\times \mathbb{H}^2$}\label{conformal_mapping}
A CFT on $\mathbb{S}^3_q$ can be mapped to a CFT on $\mathbb{S}^1\times \mathbb{H}^2$ by appropriate Weyl rescaling of the metric. The metric of $\mathbb{S}^3_q$ can be written with the coordinate transformation
\be
\sinh \eta = -\cot \theta
\ee
in the form
\be
\rmd s^2 = \sin^2\theta\left( \rmd\tau_E^2 + \ell^2(\rmd\eta^2 +  \sinh^2\eta \rmd\phi^2)\right) \ ,
\ee
where we define
\be
\tau_E = q \tau\ell ,\quad \tau_E\in [0, 2\pi q\ell)\ .
\ee
By dropping the overall Weyl scale factor  $\sin^2\theta$, we get the metric on $\mathbb{S}^1\times \mathbb{H}^2$
\be\el{StimesH2}
\rmd s^2 = \rmd\tau_E^2 + \ell^2 (\rmd\eta^2 +  \sinh^2\eta \rmd\phi^2)\ .
\ee
Under the coordinate transformation and the conformal mapping,  the North Pole $\theta = 0$ is mapped to the boundary of the hyperbolic space, $\eta \rightarrow - \infty$.

Due to the conformal nature of the CFTs, the partition function is invariant under the Weyl rescaling
\be
\el{S3=SH2}
Z[\mathbb{S}^3_q]=Z[\mathbb{S}^1_q\times \mathbb{H}^2]\ .
\ee
The background gauge field $A$ on $\mathbb{S}^3_q$ is also invariant under the Weyl rescaling\footnote{Weyl rescaling only affects the metric.}. This equality (\ref{S3=SH2}) allows us to compute partition function (and R\'enyi entropy) on a branched sphere by studying the thermal partition function on $\mathbb{S}^1\times \mathbb{H}^2$. In the case of strongly coupled CFTs, the R\'enyi entropy can be related to the thermal entropy of the dual AdS black hole \cite{Hung:2011nu} \cite{Belin:2013uta}. The conformal mapping for a free field theory can be found in \cite{Casini:2010kt} \cite{Klebanov:2011uf}. The conformal mapping also allows \cite{Casini:2011kv} to identify R\'enyi entropy of a general CFT on $\mathbb{S}^3_q$ with the R\'enyi entropy across an entangling circle in flat space.

\vskip0.75cm
\begin{figure}[ht]
\centering
   \includegraphics[scale=0.9]{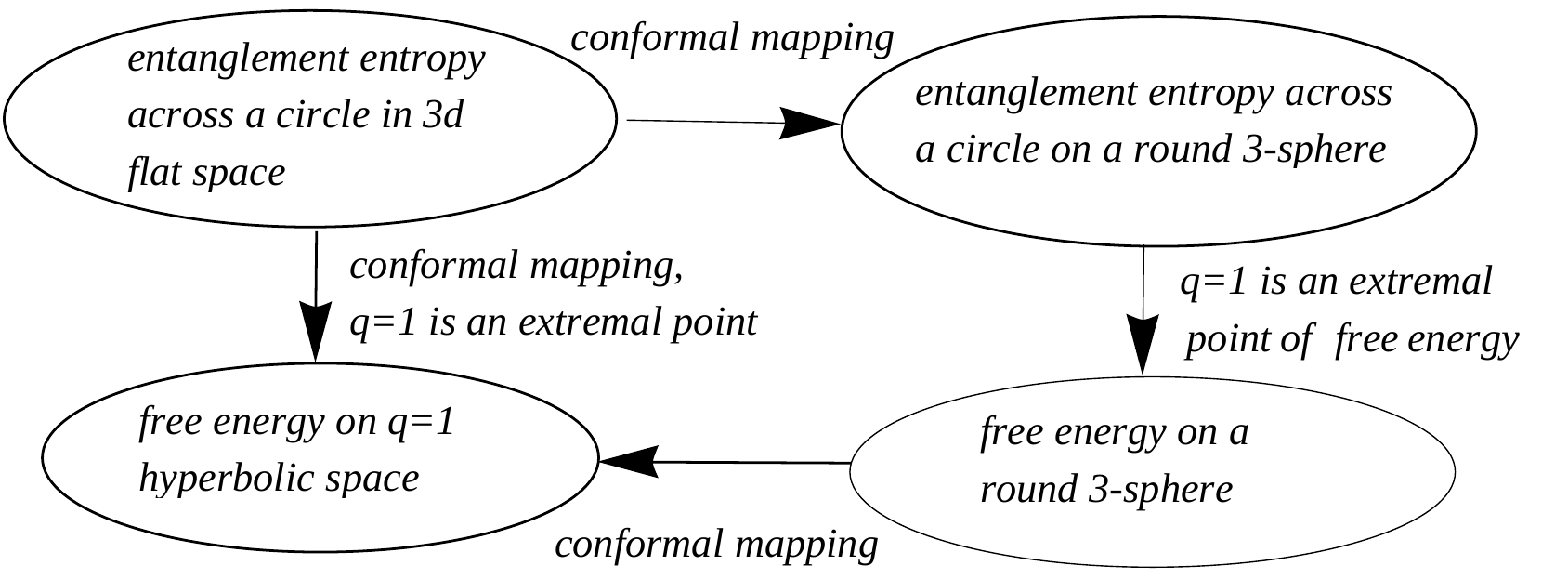}%
 \vskip0.3cm
\caption{\sl By the conformal mapping,  equivalence among the following four quantities can be established for a general CFT: (1) entanglement entropy across a circle in flat space and on a round sphere, (2) free energies on a round sphere and on a hyperbolic space. Here, $q$ is a branched parameter, which is introduced as a virtual deformation in computing the entanglement entropy by replica method.} \label{fig1}
\end{figure}
\vskip0.75cm

The coordinate transformation and the conformal mapping do not affect the supersymmetry. This is as it should be since the partition function depends only on the data of Reeb vector. Therefore, in this case, the `thermal' partition function on $\mathbb{S}^1\times \mathbb{H}^2$ is actually populated on supersymmetric ground state. In the next section, we will see this from the viewpoint of holographic dual black hole.

In Figure~\ref{fig1}, we summarize relations among these quantities.  

A remark is in order. A quantum field theory on a hyperbolic space is completely specified only if boundary conditions are specified at infinity. The situation is analogous to a quantum field theory defined on AdS space. Since boundary conditions pick ground state, different boundary condition corresponds to different specification of the quantum field theory. Therefore, conventionally, the partition function on a hyperbolic space is a function of the boundary condition. In the exact result of the partition function over the compact 3-space we studied above, the scalar field $\sigma_0$ at the North Pole (or any other point of the base space) takes a constant real value, integrated over the Coulomb branch. When conformally mapped, the image of the North Pole on the (branched) sphere is the boundary of the hyperbolic space $\mathbb{H}_2$. This implies the following things. First, the boundary condition at the boundary of $\mathbb{H}^2$ must allow an arbitrary constant value for the scalar field. Second, by the partition function on the hyperbolic space, we actually mean the integral over the boundary condition taking values in the Coulomb branch of the original, compact 3-space. Therefore, what we actually mean by (\ref{S3=SH2}) is
\be
\el{Z-definition}
Z[\mathbb{S}^3_q] = \int_{\rm Coulomb \ Branch} \rmd \sigma_0 \ Z[\mathbb{S}^1_q \times \mathbb{H}^2; \sigma_0].
\ee
Note that, by the integration domain over the Coulomb branch, we do not mean that the quantum field theory has a Coulomb branch when the theory is put on hyperbolic space. Rather, we mean that the domain of integration coincides with the Coulomb branch when the theory is put on a compact 3-space.

\section{Charged Topological Black Hole}
\label{sec:TBH}
Having established exact results on ${\cal N} = 2$ qSCFTs on branched sphere and its simplification in the large $N$ limit, we next move to establish holographic dual of these theories. In this section, we review the basics of charged topological black hole in AdS$_4$, which can be seen as the gravity dual of the thermal density matrix of a SCFT on $\mathbb{R}^1\times \mathbb{H}^2$. In the next section we will identify that the holographic dual of qSCFT is a supersymmetric charged topological black hole in AdS$_4$.

Consider an ${\cal N} =2$ SCFT on $\mathbb{R}^1\times \mathbb{H}^2$, the Lorentzian counterpart of the Euclidean SCFTs we studied in the previous section. We shall first proceed in Lorentzian signature and change to Euclidean signature in the end of this section, with the assumption that the Wick rotation can act freely in the SCFT side as well. By the AdS/CFT correspondence, one expects it to be dual to an AdS$_4$ black hole \cite{mann,brillet,vanzo,roberto,birm} with the metric\footnote{This statement is based on the assumption that $\cal N\geq$ $2$ Chern-Simons-Matter theories on $\mathbb{S}^3$ have AdS$_4$ duals \cite{Martelli:2011qj, Cheon:2011vi, Jafferis:2011zi}, with the coupling dependence of free energy encoded in the Newton's constant of AdS$_4$ gravity. We work in a general setup ---  M theory solution with the AdS$_4$ background will depend on specific SCFT under consideration. We also assume that Weyl rescaling of dual SCFTs does not break the nature of the duality. This implies that, once a given SCFT is deformed, the holography becomes more involved. }
\be
\rmd s^2 = -f(r)\rmd t^2 +{1\over f(r)}\rmd r^2 + r^2 \rmd\Sigma(\mathbb{H}^2)\label{dual_metric}\ , \ee
whose horizon is conformal to $\mathbb{R}^1 \times \mathbb{H}^2$. Here, $\rmd\Sigma(\mathbb{H}^2)$ is the conformal class metric of the intersection $\mathbb{H}^2$ of the horizon with the Cauchy surface normal to $\rmd t$, $\rmd\Sigma(\mathbb{H}^2)=\rmd\eta^2 +  \sinh^2\eta \rmd\phi^2$.
Solutions of the form (\ref{dual_metric}) was known~\cite{Romans:1991nq} in the context of four-dimensional $\cal N$ $=2$ gauged supergravity \cite{Freedman:1976aw}. Its field equations coincide with the field equations of the Einstein-Maxwell theory with negative cosmological constant\footnote{The cosmological constant is fixed by the 4d ${\cal N}=2$ supersymmetry.} \be\Lambda =-3g^2\ ,\ee where $g$ is the coupling between gauge field and gravitini.
The effective action is \footnote{This corresponds to $\ell_* =(2/g) =2
L$ in the convention of~\cite{Belin:2013uta}.
}
\be
I =-{1\over 2\ell_p^2} \int \rmd^4x \sqrt{-g} \left(2\Lambda + R - {1\over g^2} F_{\mu\nu}F^{\mu\nu}\right)\label{EMaction}\ .
\ee
Due to the relation $\Lambda = -3g^2$, the AdS radius is
\be
L={1 \over g} \ .
\ee
The factor $1\over g^2$ in front of the Maxwell Lagrangian $F^2$ can be absorbed into the definition of gauge field, the convention we will adopt from now on.
The general solution of (\ref{dual_metric}) for the action (\ref{EMaction}) is given by
\be \el{metric-f}
f(r) = {r^2\over L^2} +\kappa - {2m\over r} + {Q^2\over r^2}\ ,
\ee
where 2$\kappa$ refers to the constant curvature of two-dimensional Riemann surface. In our convention, $\kappa =-1$ for $\mathbb{H}^2$. For later convenience, we leave the value of $\kappa$ unspecified \footnote{The solution with $\kappa=-1$ discussed here has a pairing solution with $\kappa=+1$, where $\mathbb{H}^2$ is replaced by $\mathbb{S}^2$.}. The solution of the gauge field reads
\be \el{gaugefield}
A_{\zt{TBH}} = \lb {Q\over r}-\mu \rb \rmd t,
\ee
where $\mu$ is fixed by the boundary condition that the gauge field vanishes at the horizon:
\be \el{Q-mu}
\mu = {Q\over r_\text{h}}\ .
\ee
The horizon radius of black hole $r_{\text{h}}$ is given by the maximum root of the equation
\be\label{horizon_condition}
f(r_\text{h}) = 0\ ,
\ee
while the black hole temperature is determined by requiring the absence of singularity when $r\rightarrow r_\text{h}$:
\be \el{temperature}
T = {f'(r_\text{h})\over 4 \pi}\ .
\ee

\subsection{Supersymmetry}
Let's first work out the condition that the above black hole is a supersymmetric configuration.
The Killing spinor equation of the four-dimensional, $\cal N$$=2$ gauged supergravity reads
\be
\hat\nabla_\mu \epsilon = 0\label{killing_eq}\ ,
\ee
where the supercovariant derivative is given by \footnote{We use the convention $\xg_{\mu\nu} = \frac 1 2 [\xg_\mu,
\xg_\nu]$, and the 4d gamma matrices we use in the paper are listed in \er{gammasA}.}
\be
\hat\nabla_\mu = \nabla_\mu - i g A_\mu +{1\over 2}g\gamma_\mu +{i\over 4}F_{\nu\rho}\gamma^{\nu\rho}\gamma_\mu\ .
\ee
The integrability condition for (\ref{killing_eq}) leads to
\be
\Omega_{\mu\nu}\epsilon =0\label{integrability_con}\ ,
\ee
where $\Omega_{\mu\nu}$ is the tensor-spinor operator defined by
as
\ba \Omega_{\mu\nu} &:=& [\hat\nabla_\mu, \hat\nabla_\nu] \\
&=& {1\over 4}C_{\mu\nu}{}^{\rho\tau}\gamma_{\rho\tau} + {i\over 2}\gamma^{\rho\tau}\gamma_{[\nu}(\nabla_{\mu]} F_{\rho\tau}) + {i\over 8}g F_{\rho\tau}(3\gamma^{\rho\tau}\gamma_{\mu\nu}+\gamma_{\mu\nu}\gamma^{\rho\tau})\ ,
\ea where $C_{\mu\nu\rho\tau}$ is the Weyl tensor and $F_{\mu\nu}$ is the field strength. An important feature is that this operator $\Omega_{\mu\nu}$ can be factorized into the product of a nonsingular factor $X_{\mu\nu}$ and a spinor function $\Theta:$\footnote{We checked that this holds for two-dimensional Riemann surfaces, $\mathbb{S}^2$ and $\mathbb{H}^2$. We focus on the latter while the former was explicitly analyzed in~\cite{Romans:1991nq}.}
\be
\Theta := \sqrt{f(r)} + g r \gamma_1 + \left({1\over r } - {m\over Q^2}\right)i \gamma_0 Q\ .
\ee
The condition for (\ref{integrability_con}) to admit a nontrivial solution is simply the requirement of vanishing determinant of $\Theta$:
\be
\det\Theta = \frac{\left(m^2-\kappa  Q^2\right)^2}{Q^4} = 0\ .
\ee
We see that the requirement of supersymmetry condition relates the mass and the charge of the black hole:
\be\label{bps}
m^2=\kappa  Q^2.
\ee

\subsubsection{Neutral black hole}
The solution with $Q=m=0$ is a quotient space of pure AdS$_4$, describing an uncharged black hole. \footnote{The spacetime can be considered as AdS$_4$, viewed by a uniformly accelerated observer. The non-compact horizon is nothing but the observer's acceleration horizon~\cite{Romans:1991nq}.}
In this case, we have
\be
f(r) = {r^2\over L^2}-1\ ,
\ee
and the horizon radius is
\be r_\text{h}=L\ .\ee
The black hole temperature is given by
\be T_0= {1\over 2\pi L}\ .\ee
The solution is expected to be dual to a SCFT on $\mathbb{S}^1\times \mathbb{H}^2$ with $q=1$
\be
\rmd s^2 =\rmd\tau_E^2 + \ell^2 (\rmd\eta^2 +  \sinh^2\eta \rmd\phi^2)\ ,\quad \tau_E\in[0, 2\pi \ell)\ ,
\ee
which can be mapped to a CFT on a round 3-sphere, $\mathbb{S}^3$.
By matching the temperature of CFT on $\mathbb{S}^1\times \mathbb{H}^2$ and that of the black hole, the AdS$_4$ radius is set to be the size of the 3-sphere
\be
L = \ell\ .
\ee
Killing spinor equation (\ref{killing_eq}) with $m=Q=0$ has nontrivial solutions and the geometry is at least locally supersymmetric~\cite{Caldarelli:1998hg}.
This is consistent with the fact that, for $q=1$, the field theory is supersymmetric with no additional background gauge field. We will not come to the details of the Killing spinor solutions since the uncharged topological black hole is not our main focus. Notice that, in the bulk, the hyperbolic horizon with $q=1$ can be mapped to Ryu-Takayanagi surface~\cite{Ryu:2006bv} with the mapping between boundaries discussed at the end of section~\ref{conformal_mapping}.

\subsubsection{Charged black hole}
For the solution with $Q^2=\kappa m^2 \neq 0$, $f(r)$ in the metric takes the form
\be
f(r)={r^2\over L^2}+\kappa \left(1-{m\over \kappa r}\right)^2\ ,
\ee
where we used $|\kappa|=1$. The Killing spinor equation \er{killing_eq} can be solved following~\cite{Romans:1991nq}, and we will do so in section~\ref{subsubsection:Killing spinor}. As we shall see later, in solving \er{killing_eq} it is very helpful to use the integrability condition
\be
\Theta \epsilon = 0\ , \quad\Theta = \sqrt{f(r)} + g r \gamma_1 + \left({1\over r } - {1\over \kappa m}\right)i \gamma_0 Q\ ,
\ee
and construct a projection operator
\be
P := { \Theta\over 2\sqrt{f(r)}}\ .
\ee

\section{TBH$_4$/qSCFT$_3$ Correspondence}
\label{sec:TBH/qSCFT}
In this section, we would like to show that the three-dimensional ${\cal N}=2$ Chern-Simons-Matter theory on a $q$-branched sphere ($q>1$) is holographically dual to the supersymmetric charged topological black hole. To support the duality, we compute free energy and R\'enyi entropy from the topological black hole (following the approach in \cite{Belin:2013uta}) and find that they agree perfectly with the results from the qSCFT. We also show that four-dimensional Killing spinors are reduced to the three-dimensional Killing spinors at the boundary $\mathbb{S}^1\times \mathbb{H}^2$. 

\subsection{Free energy and R\'enyi entropy}
We shall first compute the R\'enyi entropy holographically from the charged topological black hole specified by metric (\ref{dual_metric}) and gauge field (\ref{gaugefield}). This can be done by studying the thermodynamics of the black hole. We work in grand canonical ensemble. In this ensemble, the Gibbs potential is given by
\be
W = I/\beta = E-TS-\mu \widehat Q\ ,
\ee
where $I$ is the Euclidean on-shell action and $\beta=1/T$ denotes the period of Euclidean time direction $\tau_E$. The state variables can be computed as follows:
\ba
E &=& \left(\partial I \over \partial \beta \right)_\mu - {\mu\over \beta}\left(\partial I \over \partial \mu\right)_\beta\ ,\\
S &=& \beta  \left(\partial I \over \partial \beta \right)_\mu - I\label{entropy1}\ ,\\
\hat Q &=& -{1\over \beta}\left(\partial I \over \partial \mu\right)_\beta\label{totalcharge}\ .
\ea
Let's consider the black hole with both finite charge and temperature. The free energy is given by
\be
I:= \log Z(\mu, T)\ .
\ee
Here, both $\mu$ and $T$ are functions of parameter $q$ only. This follows because temperature of the black hole is fixed by matching it to that of the boundary CFT on $\mathbb{S}^1\times \mathbb{H}^2$
\be\label{temperature1}
T(q) = T_0/q\ ,
\ee
while chemical potential $\mu$ is fixed by matching it to the background gauge field of the boundary SCFT
\be\label{chemical_potential}
\mu(q) =-\left({q-1\over 2q}\right)i\ .
\ee
We now compute R\'enyi entropy defined in eq.\er{SReyni}. It can be written as an integral over branched parameter $n$
\be\label{integralovern}
S_q = {q\over q-1}\left({\log Z_1\over 1}-{\log Z_q \over q}\right) = {q\over q-1}\int_q^1 \partial_n\left({\log Z(T, \mu)\over n}\right) \rmd n\ .
\ee
By using (\ref{entropy1}), (\ref{totalcharge}) and (\ref{temperature1}), the total derivative term in (\ref{integralovern}) can be written as
\be
\partial_q\left({\log Z(T, \mu)\over q}\right) = {S\over q^2} - {\widehat Q \mu'(q)\over T_0}\ ,
\ee
where $\widehat Q$ is the total charge of black hole. The charge $\widehat Q$ can be computed from the Gauss's law:
\be
\widehat Q = {2V_\Sigma\over \ell_p^2}Q =  \left({2V_\Sigma \over \ell_p^2}\right)\mu(q) r_\text{h}\ ,
\ee
where $V_\Sigma$ denotes the volume of $\mathbb{H}^2$. The thermal entropy is given by the Bekenstein-Hawking formula
\be
S_{\rm BH} =2\pi {V_\Sigma\over \ell_p^2} r_{\text{h}}^2\ .
\ee
The horizon radius $r_\text{h}$ can also be expressed as a function of $q$ by combining (\ref{horizon_condition}) and (\ref{temperature}) and substituting in (\ref{temperature1}) and (\ref{chemical_potential})
\be\label{horizon_q}
x(q):={r_\text{h}\over L} =\frac{1}{3q} \Big[\sqrt{3~\mu(q) ^2  q^2+3 q^2+1}+1 \Big] \ .
\ee
Substituting $S(q)\ ,\widehat Q(q)\ ,\mu(q)$, we can finally express the integral (\ref{integralovern}) as
\ba
\begin{aligned}
S_q &=  2\pi \lb\frac {L} {\ell_p}\rb^2 V_\xS {q\over q-1}\int_q^1 \left({x(n)^2\over n^2} - 2 x(n) \mu(n) \mu'(n)\right) \rmd n\\
       &= 2\pi \lb\frac {L} {\ell_p}\rb^2 V_\xS  {q\over q-1}\int_q^1 \frac{n+1}{2 n^3} \rmd n\\
       &= \frac{3 q+1}{4 q} S_1\ .
\end{aligned}
\ea
We see that this agrees precisely with the CFT result \er{super-Reyni-ft}.

It is also straightforwardly seen that the free energy agrees between the black hole and the CFT. This follows from the same relation of partition function as (\ref{qpartition})
\be
I_q = {(q+1)^2\over 4q} I_1\ ,
\ee
which can be seen from the definition of R\'enyi entropy (\ref{SReyni}) and the known fact that $S_1=I_1$.

Actually, one can check that, at $q=1$, for general strongly coupled three-dimensional CFTs, we have a chain of identities
\ba
&& \log Z[\mathbb{S}^3] \nn &=& \text{entanglement entropy across $\mathbb{S}^1$ on $\mathbb{S}^3$} \nn
&=& \text{entanglement entropy across $\mathbb{S}^1$ on $\mathbb{R}^{1,2}$} \nn
&=& I[\text{AdS}_4] \nn
&=& \text{Ryu-Takayanagi Entanglement Entropy}\nn
&=& \log Z[\mathbb{S}^1\times\mathbb{H}^2] \nn
&=&\text{Bekenstein-Hawking entropy} \ S_{\rm BH} [\zt{TBH}_4] \nn
&=& I[\zt{TBH}_4]\ .
\ea
Once again, by the free energy on $\mathbb{S}^1 \times \mathbb{H}^2$, we mean log of the partition function defined by the integral over the Coulomb branch, as in (\ref{Z-definition}).

Notice that the background gauge field on $\mathbb{S}^3_q$ given by \er{backgroundfield}
implies an imaginary $\mu$ (by eq.(\ref{chemical_potential})). We see this follows from the relation
\be g A_{\zt{TBH}} (r\to \infty)= A(\mathbb{S}^3_q)\ ,
\ee
and $t=-i \tau_E$.

\subsection{Supersymmetry}
If the TBH$_4$ is holographically dual to qSCFT, it must be preserving two supercharges of opposite R-charges. We will now show that the TBH$_4$ with chemical potential \er{chemical_potential} and temperature \er{temperature1} is in fact supersymmetric. We will also show that the Killing spinors obey the holographic relations -- when restricted to the boundary, the four-dimensional Killing spinors are reduced to those on $\mathbb{S}^1\times \mathbb{H}^2$ at radial infinity, up to conformal rescaling.

\subsubsection{Mass-Charge Relation}
We first check the mass and charge relation for the topological AdS$_4$ black hole we are considering.
The mass parameter can be solved from \er{horizon_condition}, in terms of $x$ and $Q$:
\be m = \frac 1 2 \lb x (x^2 -1)L + \frac {Q^2} {x L} \rb\ .\ee
Substituting the chemical potential \er{chemical_potential} back into \er{horizon_q}, the horizon radius (in unit of $\ell$) can be simplified to
\be x(q) = {1\over 2}\left(1+\frac {1}{q}\right)\ .\ee
Substituting chemical potential \er{chemical_potential} into $Q-\mu$ relation \er{Q-mu} and using the simplified $x(q)$, we have
\be Q(q) = -{i\over 4}L\left(1-\frac { 1}{q^2}\right)\ . \ee
Finally, the mass can be expressed as a function of $q$
\be
m(q) =-{1\over 4}L\left(1-\frac { 1}{q^2}\right)\ .
\ee
Therefore, the supersymmetry condition \er{bps}  is satisfied with $Q=i\, m$
\be
m^2+Q^2 = 0\ .
\ee
As discussed in Sec~\ref{sec:TBH}, this implies that the integrability condition is satisfied. We then expect to find nontrivial solutions to the Killing spinor equations \er{killing_eq}, which we will do in the next subsection.

\subsubsection{Killing spinor}
\label{subsubsection:Killing spinor}
Let's now explicitly solve for Killing spinors on TBH$_4$ with the boundary metric \er{StimesH2}
\be
\rmd s^2 = -f(r)\rmd t^2 +{1\over f(r)}\rmd r^2 +  r^2 (\rmd\eta^2 +  \sinh^2\eta \rmd\phi^2)\label{E-TBH4}\ , \ee
where $f(r)$ is given by \er{metric-f}  with $\xk = -1, Q=i\, m $
\be
f(r) = \frac{r^2}{L^2}- \left(1+{m\over r}\right)^2\ .
\ee
The vielbeins are
\ba e^0 =\sqrt{f(r)}\rmd t, &\quad &e^1 = \frac {\rmd r}{\sqrt{f(r)}}, \nn
e^2 = r \rmd \eta,&\quad &e^3 = r \sinh \eta \rmd \phi,
\ea
and the nonvanishing components of the spin connection are
\ba
\xo_{t}{}^0{}_1 = \frac 1 2 f'(r),\quad & & \xo_{\eta}{}^1{}_2 = -\sqrt{f(r)},\nn
\xo_{\phi}{}^1{}_3 = -\sqrt{f(r)}\sinh\eta,\quad & & \xo_{\phi}{}^2{}_3 = -\cosh\eta\ .
\ea
Combining the spin connection and the field strength, we have the supercovariant derivatives
\ba
\begin{aligned}
\el{supercovariantd}
\hat{\nabla}_{t} &= \partial_{t} - i{1\over L}\lb \frac{Q}{r} -\frac{Q}{r_\text{h}}\rb+ \frac{1}{2L}
                   \sqrt{f(r)}\xg_0 - i\frac{Q }{2 r^2}\sqrt{f(r)}\xg_1
                   +{1\over 4}f'(r)\xg_{01}, \\
\hat{\nabla}_r &= \partial_r + \frac{1}{2L}\sqrt{f(r)}^{-1}\xg_1 - i\frac{Q }{2 r^2}\sqrt{f(r)}^{-1}\xg_0, \\
\hat{\nabla}_{\eta} &= \partial_{\eta} - \frac{1}{2}\sqrt{f(r)}\xg_{12} +
                          \frac{r}{2L}\xg_2- i\frac{Q }{2 r}\xg_{01}\xg_2, \\
\hat{\nabla}_{\phi} &= \partial_{\phi} - \frac{1}{2}\sqrt{f(r)}\xg_{13}
                        \sinh\eta - \frac{1}{2}\xg_{23}\cosh\eta
                        + \frac{1}{2L}
                        r\xg_3\sinh\eta  - i\frac{Q }{2 r}\sinh\eta\xg_{01}\xg_3 \ .
\end{aligned}
\ea
The projection operator $P$ is defined as
\be
P :={\Theta \over 2\sqrt{f(r)}}= \frac 1 2 \lb 1 -\frac {1}{\sqrt{f(r)}}\lb 1 + \frac m r\rb\xg_0 + \frac {1}{\sqrt{f(r)}}\frac r L\xg_1\rb\ .
\ee
We can use the integrability condition $P \xe = 0$ to simplify \er{supercovariantd}. The Killing spinor equations \er{killing_eq}
can be finally expressed as
\ba
\lb \p_t - {1\over 2L}\left(1+{2 m/ r_\text{h}}\right) \rb \xe & = & 0 \el{temporal}\\
\lb \p_r + \frac m{2r( r + m)}+\frac 1 {2L\sqrt{f(r)}}(1+\frac {m} {r+ m}) \xg_1\rb \xe & = & 0 \el{radial}\\
\lb \p_\eta - \frac 1 {2} \xg_0\xg_1 \xg_2 \rb \xe & = & 0 \el{angular1}\\
\lb \p_\phi - \frac 1 {2} \cosh\eta \xg_{23} -\frac 1 2 \sinh\eta(\xg_0\xg_1\xg_3) \rb \xe & = & 0 \el{angular2}
\ea
This type of equations can be solved \cite{Romans:1991nq}. All of the supercovariant derivatives commute with each other except for $\hat \cd_\eta$ and $\hat \cd_\phi$. We can solve the radial, temporal and angular equations separately. $t\ ,\eta\ ,\phi$ components are solved first. The solution can be expressed as
\be \el{4dsol} \xe(t, r, \eta, \phi) = e^{\frac 1 {2 q L} t} e^{\frac \eta 2 \xg_0\xg_1\xg_2} e^{\frac \phi 2 \xg_{23}} \xe(r).\ee
The radial equation takes the form of
\[\p_r \xe(r) = (a(r)+b(r)\xG_1)\xe(r),\]
and $\xe(r)$ also satisfies the constraint $P\xe(r) = 0$ with $P$ in the form of
\[P = \frac 1 2 (1+ x(r)\xG_1+y(r)\xG_2)\ ,\]
where $\xG_{1,2}$ are matrices satisfying
\be \xG_1^2 = \xG_2^2 = 1, \quad \xG_1 \xG_2
+ \xG_2 \xG_1 = 0\ .\ee
Solution to this type of equation is provided in the appendix of \cite{Romans:1991nq}
\be \el{solspin}
\xe(r) = (u(r)+v(r)\xG_2)\lb \frac {1-\xG_1} 2 \rb\xe_0,
\ee
where $u$, $v$ are defined by,
\be u = \sqrt{1 + x\over y} e^w,\quad v =- \sqrt{1 - x\over y} e^w,\quad
w(r) = \int^r a(r') dr',
\ee
and $\xe_0$ is an arbitrary constant spinor.
In our case, \er{solspin} gives
\be \el{radialsol}\xe(r) = \lb \sqrt{\frac r L + \sqrt{f(r)}}-\xg_0 \sqrt{\frac r L - \sqrt{f(r)}}\rb \lb \frac {1-\xg_1} 2 \rb \xe'_0\ , \ee
where $\xe'_0$ is an arbitrary constant spinor.

Similarly, the Killing spinor in the Euclidean TBH$_4$ background is given by
\be \el{4dsolE} \xe(\tau_E, r, \eta, \phi) = e^{-\frac i {2 q L} \tau_E} e^{i\frac \eta 2 \xg_0\xg_1\xg_2} e^{\frac \phi 2 \xg_{23}} \xe(r),\ee
with
\be \el{radialsoE}\xe(r) = \lb \sqrt{\frac r L + \sqrt{f(r)}}-i\xg_0 \sqrt{\frac r L - \sqrt{f(r)}}\rb \lb \frac {1-\xg_1} 2 \rb \xe'_0\ . \ee

\subsubsection{Holography of Killing spinors}
The pre-requisite of the holographic relation we proposed above is that the Killing spinors in the background of THB$_4$ must reduce to the Killing spinors on branched 3-sphere the qSCFT$_3$ is defined. Here, we will check this  by showing that the Killing spinor equations on TBH$_4$ is reduced at asymptotic infinity to the Killing spinor equation on branched 3-sphere, up to conformal rescaling. Hereafter, we take the convention of Dirac gamma matrices listed in \er{gammasA}. Notice that the projection operator $(1-\gamma_1)/2$ will project out the second and fourth components for a 4-spinor
\[
\text{$ {1 \over 2} \Big[1-\xg_1 \Big] \xe_0$}=\left(
\begin{array}{c}
 a \\
 0 \\
 c \\
 0 \\
\end{array}
\right)
\ .\]
We can temporarily drop the $r$ dependent factor. Then, the Killing spinor \er{4dsol} becomes
\be
\epsilon' = e^{\frac 1 {2 q L} t} e^{\frac \eta 2 \xg_0\xg_1\xg_2} e^{\frac \phi 2 \xg_{23}} \frac {1-\xg_1} 2 \xe_0\ ,
\ee
which can be evaluated to be
\[\epsilon' = e^{\frac 1 {2 q L} t} \left(
\begin{array}{c}
 M \\
 0 \\
 N  \\
 0 \\
\end{array}
\right),\]
where
\be
M= a\,e^{\frac{\eta}{2}} \cos \left(\frac{\phi }{2}\right)-c\,e^{\frac{\eta}{2}} \sin \left(\frac{\phi }{2}\right)\ ,\quad N=c\, e^{-\eta/2} \cos \left(\frac{\phi }{2}\right)+a\, e^{-\eta/2} \sin \left(\frac{\phi }{2}\right)\ .
\ee
Indeed, the solution contains the first and the third components only. This indicates that the 4-component spinor equations is  decomposable such that only $a(t,\eta,\phi)$ and $c(t,\eta,\phi)$ components are left out. It is convenient to start from the simplified Killing spinor equations
\er{temporal}\er{angular1}\er{angular2}. Notice that the kinetic operators in $t$ and $\eta$ components are diagonal and therefore the reduction is straightforward. For $\phi$ component, the matrix after the derivative can be written as
\[
\left(
\begin{array}{cccc}
 0 & 0 & L^1_\phi & 0 \\
 0 & 0 & 0 & L^2_\phi \\
 L^2_\phi & 0 & 0 & 0 \\
 0 & L^1_\phi & 0 & 0 \\
\end{array}
\right)
\]
where
\be
L^1_\phi = \frac{\cosh \eta}{2}+\frac{\sinh \eta}{2}\ ,\quad L^2_\phi = -\frac{\cosh \eta}{2}+\frac{\sinh \eta}{2}\ .
\ee
We see that the reduced 2-component spinor equation is given by
\be
\left(\partial_\phi + {i\cosh \eta\over 2}\sigma_2 +{\sinh \eta\over 2}\sigma_1 \right)\epsilon= 0\ ,
\ee where the 2-spinor $\epsilon$ is defined as
\be
\epsilon:= \left(
\begin{array}{c}
 a(t,\eta,\phi) \\
 c(t,\eta,\phi)\\
\end{array}
\right)\ .
\ee
Let's further perform the Wick rotation $t\to -i\tau_E$. Then, the $\tau_E$ and $\eta$ components become
\ba
\left(\p_\eta - \frac 1 {2} \sigma_3 \right) \xe & =&  0\ ,\\
\lb \p_{\tau_E} + {i\over 2q\ell} \rb \xe & = & 0\ .
\ea
We recognize that these equations are identifiable with the three-dimensional Killing spinor equations on $\mathbb{S}^1\times \mathbb{H}^2$:
\be\label{kses1h2}
\left(\nabla_\mu - i A_\mu +{i\over 2\ell}e^{\bar\nu}_\mu \gamma_{\bar\nu}\gamma_{\bar
\tau_E}\right)\epsilon = 0\ .
\ee
Here,  $\bar\nu$ denotes the flat indices and the three-dimensional Dirac gamma matrices are defined by Pauli matrices
\be
\gamma_{\bar\tau_E} =\sigma_2\ ,\quad \gamma_{\bar{\eta}} = \sigma_1\ ,\quad \gamma_{\bar \phi} = \sigma_3\ ,
\ee
Moreover, the background gauge field $A={1\over 2}\left(q-1\right)\rmd\tau $ is precisely the one we had to turn on over branched 3-spheres to preserve the two supercharges of opposite R-charges. (\ref{kses1h2}) will be connected to (\ref{kseq1})(\ref{kseq2}) by conformal rescaling and coordinate transformation.

The reduction of the Killing spinor equations implies that one can solve for the Killing spinors explicitly. The three-dimensional Killing spinors at radial infinity set boundary condition of the spinors for the four-dimensional Killing spinor equations. Up to conformal rescaling, one gets nontrivial Killing spinors on TBH$_4$ from nontrivial Killing spinors on branched 3-sphere. We conclude that the holographic relation of Killing spinors is injective.

%
%
%
%
\section{Conclusion and Discussions}
\label{sec:conclusion}
In this work, we studied three-dimensional ${\cal N}=2$ supersymmetric field theories on a general class of branched 3-spheres with $U(1)\times U(1)$ isometry. We showed that supersymmetry localization techniques can be made to work even on these singular spaces. We have particularly shown that all the branched spheres belonging to this class have the same form of Reeb vectors and therefore have the same form of partition functions as a function of deformation parameter. We focused on $\mathbb{S}^3_q$ as the representative for this class of spaces and found that there is a natural gravity dual for SCFT on it. As supporting evidences, we computed the holographic free energy and R\'enyi entropy and confirmed that they precisely agree with the large $N$ result of corresponding exact results of SCFT$_3$ by the supersymmetry localization techniques. This agreement also indicates that the supersymmetry localization techniques can be utilized on singular spaces such as branched 3-spaces. Built upon these facts, we proposed TBH$_4$/qSCFT$_3$ correspondence, which is the correspondence between SCFT on $q$-branched sphere and supersymmetry-preserving, charged topological black hole in AdS$_4$. We checked the proposed correspondence by matching free energy, R\'enyi entropy and supersymmetries. We believe further checks can be made for other physical observables such as supersymmetric Wilson loops and correlation functions.

It was recently understood that three-dimensional partition functions with ${\cal N}=2$ supersymmetry depends only on the almost contact metric structure of the three-dimensional manifold. Specifically $Z_{{\cal M}_3}$ (of $\CM_3$ with $U(1) \times U(1)$ isometry) only depends on the Reeb vector.
It would be interesting to understand this result entirely in terms of the topological black hole in AdS$_4$. Note that $q$ dependence in the Reeb vector on the branched 3-sphere is the same as the $q$ dependence in the temperature $T$ of the topological black hole. We thus expect that the Reeb vector dependence of $Z_{{\cal M}_3}$ is mapped to the $T$ dependence of the black hole partition function.

To make further check in this direction, we can consider the resolution deformation of the black hole geometry. It is clear that, adding the resolution function (\ref{eq:resolved space}) corresponds to adding a factor $R_\epsilon(\eta)$ in the line element of $\rmd\eta^2$. Therefore, the black hole metric becomes
\be\label{resolvedBH}
\rmd s^2 = -f_\epsilon(r,\eta)\rmd t^2 + g_\epsilon(r,\eta)\rmd r^2 + r^2\biggr[(1+R_\epsilon(\eta)) \rmd\eta^2 +  \sinh^2\eta \rmd\phi^2\biggr]\ .
\ee
Since the singularity in the branched 3-sphere $\mathbb{S}^3_q$ is mapped to the boundary at radial infinity of $\mathbb{H}^2$, resolving the singularity will correspond to small deformation of $\mathbb{H}^2$.
We expect that the resolved black hole is still a solution of Einstein-Maxwell theory, with a flux of the gauge field depending on the resolution $\epsilon$. Nevertheless, the free energy of the black hole ought not to depend on $\epsilon$. Reverting the direction, if we relax $\epsilon$ not to be small, the above procedure would lead us to the gravity dual of a SCFT$_3$ on an ellipsoid, which is a regular 3-manifold. Again, the free energy of the black hole will not change. Provided such gravity dual (\ref{resolvedBH}) exists, we may claim that the supersymmetry preserving condition makes the partition function solely depend on temperature $T$, equivalently, a single parameter $q$ since $T=T_0/q$ is independent of the resolution. This is in fact the black hole version of the same statement in the SCFT that global supersymmetry will restrict $Z_{{\cal M}_3}$ to be a function of a single complex-valued deformation parameter $\gamma$ in (\ref{xTeta}) (or equivalently $b$ in \er{bpara}).


It is also interesting to compare our solution with the one found in~\cite{Martelli:2011fu}. This can be done by performing a coordinate transformation in the bulk (or a different slicing in the language of \cite{Emparan:1999pm}) so that the boundary of the TBH (or its resolved version) becomes a (resolved) $q$-branched $\mathbb{S}^3$. Note that despite the conical singularity in the boundary theory, the gravity solution is smooth everywhere \cite{Lewkowycz:2013nqa}.

\section*{Acknowledgement}
We are grateful for useful discussions with SNU String Theory group members and with Zohar Komargodski, Dario Martelli and Adam Schwimmer. This work was supported by the National Research Foundation of Korea(NRF) grant funded by the Korea government(MSIP) through the Center for Quantum Spacetime(CQUeST) of Sogang University with grant number 2005-0049409 (SJR, YZ), and  through Seoul National University with grant numbers 2005-0093843, 2010-220-C00003 and 2012K2A1A9055280 (XH, SJR).

\appendix

\section{Notations and Conventions}
\subsection{Three Dimensions}
Consider a 3-dimensional spin manifold ${\cal M}_3$. In Lorentzian signature, the tangent space has the Lorentz symmetry Spin$(2,1) \simeq SU(1,1)$. A spinor transforms as a defining representation of $SU(1,1)$. On the tangent space,  we have fundamental symbols $\eta_{mn}, \eta^{mn}$, Levi-Civita antisymmetric symbol $\varepsilon_{mnp}$, and spinor antisymmetric symbol $\varepsilon_{\alpha \beta}$.
They satisfy the relations
\ba
&& \eta_{mp} \eta^{pn} = \delta^n_m \equiv \mbox{diag} (+,+,+) \nn
&& \varepsilon_{mnr} \varepsilon^{rpq} = \delta_m^p \delta_n^q - \delta_m^q \delta_n^p \nn
&& \varepsilon_{\alpha \gamma} \varepsilon^{\gamma \beta} = \delta^\beta_\alpha \equiv \mbox{diag}(+,+).
\ea
The spinors are complex-valued, and complex conjugation is an internal operation. Therefore, $\psi_\alpha$ and its complex conjugate $\psi^*_\alpha$ are related each other by charge conjugation. In Euclidean signature, the tangent space has the Lorentz symmetry Spin$(3) \simeq SU(2)$. The spinors are complex-valued, and complex conjugation is an external operation. Therefore, $\psi_\alpha$ and $\psi^*_\alpha$ are mutually independent.

In the main text, we adopted the convention of the $(2 \times 2)$ Dirac gamma matrices in tangent space to the Euclidean round three sphere as \footnote{We choose different gamma matrices for squashed sphere and $\mathbb{S}^1\times\mathbb{H}^2$ in our convention.}
\be
\gamma_1 = \sigma_1, \quad \gamma_2 = \sigma_2,\quad \gamma_3 = \sigma_3\ .
\ee
where the $\sigma_1, \sigma_2, \sigma_3$ are the Hermitian Pauli matrices.

\subsection{Four Dimensions}
On a 4-dimensional Lorentzian spin manifold, the tangent space has the Lorentz symmetry Spin$(3,1) \simeq SL(2,\BC)$. A spinor transforms as a defining representation of $\zt{SL}(2,\BC)$. On the tangent space, we have fundamental symbols $\eta_{mn}, \eta^{mn}$, Levi-Civita antisymmetric symbol $\varepsilon_{mnpq}$, and spinor antisymmetric symbol $\varepsilon_{\alpha \beta}$.
They satisfy the relations
\ba
&& \eta_{mp} \eta^{pn} = \delta^n_m \equiv \mbox{diag} (+,+,+,+) \nn
&& -\frac 1 2\varepsilon_{mnrs} \varepsilon^{pqrs} = \delta_m^p \delta_n^q - \delta_m^q \delta_n^p \nn
&& -\frac 1 6 \varepsilon_{mprs} \varepsilon^{nprs} = \delta_m^n \nn
&& \varepsilon_{\alpha \gamma} \varepsilon^{\gamma \beta} = \delta^\beta_\alpha \equiv \mbox{diag}(+,+).
\ea
The spinors are complex-valued, and complex conjugation take a spinor in one Weyl representation to its conjugate representation. In Euclidean signature, the tangent space has the Lorentz symmetry Spin$(4) \simeq SU(2) \times SU(2)$. The spinors are complex-valued, and a spinor $\psi_\alpha$ and its complex conjugate $\psi^*_\alpha$ transform under the same representations $(\df 2,\df 1)$. Therefore, chiral spinors $\psi, \widetilde \psi$ in different Weyl representations are mutually independent.

We choose the following 4d real gamma matrices in Lorentz signature,
\begin{align}
\label{gammasA}
\begin{aligned}
\gamma_0 & =\left(
\begin{array}{cccc}
 0 & 0 & 0 & -1 \\
 0 & 0 & 1 & 0 \\
 0 & -1 & 0 & 0 \\
 1 & 0 & 0 & 0 \\
\end{array}
\right)\ ,\quad
\gamma_1=\left(
\begin{array}{cccc}
 -1 & 0 & 0 & 0 \\
 0 & 1 & 0 & 0 \\
 0 & 0 & -1 & 0 \\
 0 & 0 & 0 & 1 \\
\end{array}
\right)\ , \\
\gamma_2 & =\left(
\begin{array}{cccc}
 0 & 0 & 0 & -1 \\
 0 & 0 & 1 & 0 \\
 0 & 1 & 0 & 0 \\
 -1 & 0 & 0 & 0 \\
\end{array}
\right)\ ,\quad
\gamma_3=\left(
\begin{array}{cccc}
 0 & 1 & 0 & 0 \\
 1 & 0 & 0 & 0 \\
 0 & 0 & 0 & 1 \\
 0 & 0 & 1 & 0 \\
\end{array}
\right)\ .
\end{aligned}
\end{align}

\section{Branched 3-spheres}
\subsection{Round}\label{vielbeinS3q}
We choose 3d gamma matrices in terms of Pauli matrices \be
\gamma_1=\sigma_1\ ,\quad\gamma_2=\sigma_2\ ,\quad\gamma_3=\sigma_3\ .
\ee for round sphere and ellipsoid. The vielbein for $\mathbb{S}^3_q$ is
\begin{align}
\el{vielbeinround}
\begin{aligned}e^{1}/\ell &= \mu^1 =\sin(\tau+\phi)\rmd\theta+\cos(\tau+\phi)\sin\theta\cos\theta(q \rmd\tau- \rmd\phi)\ ,\\
e^{2}/\ell & = \mu^2 =-\cos(\tau+\phi)\rmd\theta+\sin(\tau+\phi)\sin\theta\cos\theta(q \rmd\tau- \rmd\phi)\ ,\\
e^{3}/\ell & = \mu^3 =q\sin^{2}\theta \rmd\tau+\cos^{2}\theta \rmd\phi\ ,
\end{aligned}
\end{align}
with a spin connection
\begin{align}
\el{spinconnectionround}
\begin{aligned}\omega^1_{~2} & =(1-q\cos^{2}\theta)\rmd\tau+\cos^{2}\theta \rmd\phi\ ,\\
\omega^1_{~3} & =\cos(\tau+\phi)\rmd\theta-\sin\theta\cos\theta\sin(\tau+\phi)(q \rmd\tau- \rmd\phi)\ ,\\
\omega^2_{~3} & =\sin(\tau+\phi)\rmd\theta+\sin\theta\cos\theta\cos(\tau+\phi)(q \rmd\tau- \rmd\phi)\ .
\end{aligned}
\end{align}
Notice that the basis we used here are left invariant frame (corresponding to $H=-i$) .
\subsection{Ellipsoid}
We choose a vielbein
\begin{align}
\begin{aligned}e^{1} & =f(\theta)\sin(\tau+\phi)\rmd\theta+\cos(\tau+\phi)\sin\theta\cos\theta(q\tilde{\ell}\rmd\tau-p\ell \rmd\phi)\ ,\\
e^{2} & =-f(\theta)\cos(\tau+\phi)\rmd\theta+\sin(\tau+\phi)\sin\theta\cos\theta(q\tilde{\ell}\rmd\tau-p\ell \rmd\phi)\ ,\\
e^{3} & =q\tilde{\ell}\sin^{2}\theta \rmd\tau+p\ell\cos^{2}\theta \rmd\phi\ ,
\end{aligned}
\label{eq:Resolved_space_vielbein}
\end{align}
with a spin connection
\begin{align}
\begin{aligned}\omega_{~2}^{1} & =(1- {q\tilde{\ell}\over f(\theta)}\cos^{2}\theta)\rmd\tau+(1-{p\ell\over f(\theta)}\sin^{2}\theta)\rmd\phi\ ,\\
\omega_{~3}^{1} & =\cos(\tau+\phi)\rmd\theta-{1\over f(\theta)}\sin\theta\cos\theta\sin(\tau+\phi)(q\tilde \ell \rmd\tau-p\ell \rmd\phi)\ ,\label{eq:Resolved_{s}pace_{s}pin_{c}onnection}\\
\omega_{~3}^{2} & =\sin(\tau+\phi)\rmd\theta+{1\over f(\theta)}\sin\theta\cos\theta\cos(\tau+\phi)(q\tilde \ell \rmd\tau-p\ell \rmd\phi)\ .
\end{aligned}
\end{align}
\subsection{Squashed sphere}\label{branchedsquashed_vielbein}
We choose gamma matrices in terms of Pauli matrices
\be
 \gamma_1=\sigma_3\ ,\quad \gamma_2=-\sigma_1\ ,\quad\gamma_3=-\sigma_2\ .
\ee
 and vielbein ($\ell=1$)
\begin{align}
\begin{aligned}
e^{1} & =-{q\over 2v}(1+\cos2\theta)\rmd\tau - {p\over 2v}(1-\cos2\theta)\rmd\phi\ ,\\
\label{squashed_vielbein} e^{2} & =-\sin(\tau+\phi)\rmd\theta+ {q\over 2}\sin2\theta \cos(\tau+\phi)\rmd\tau - {p\over 2} \sin2\theta\cos(\tau+\phi)\rmd\phi\ ,\\
e^{3} & =\cos(\tau+\phi) \rmd\theta + {q\over 2}\sin2\theta\sin(\tau+\phi)\rmd\tau  -{p\over 2}\sin2\theta \sin(\tau+\phi)\rmd\phi\ ,
\end{aligned}
\end{align}
with a spin connection
\begin{align}
\begin{aligned}\omega_{~2}^{1} & =-{\cos(\tau+\phi)\over v}\rmd\theta -{q\over 2v}\sin2\theta\sin(\tau+\phi)\rmd\tau + {p\over 2v}\sin2\theta\sin(\tau+\phi)\rmd\phi\ ,\\
\omega_{~3}^{2} & = \left(\left(2-{1\over v^2}\right)q\cos^2\theta+1-q\right)\rmd\tau + \left(\left(2-{1\over v^2}\right)p\sin^2\theta+1-p\right)\rmd\phi\ ,\\
\omega_{~1}^{3}& ={\sin(\tau+\phi)\over v}\rmd\theta-\frac{q \sin2\theta\cos (\tau +\phi )}{2 v}\rmd\tau + \frac{p \sin2\theta\cos (\tau +\phi )}{2 v}\rmd\phi\ .
\end{aligned}
\end{align}


\begin{thebibliography}{100}

\bibitem{Pestun:2007rz}
  V.~Pestun,
  ``Localization of Gauge Theory on a Four-sphere and Supersymmetric Wilson loops,''
  Commun.\ Math.\ Phys.\  {\bf 313}, 71 (2012)
  [arXiv:0712.2824 [hep-th]].

\bibitem{Kapustin:2009kz}
  A.~Kapustin, B.~Willett and I.~Yaakov,
  ``Exact Results for Wilson Loops in Superconformal Chern-Simons Theories with Matter,''
  JHEP {\bf 1003}, 089 (2010)
  [arXiv:0909.4559 [hep-th]].

\bibitem{Kallen:2012cs}
  J.~K\"all\'en and M.~Zabzine,
  ``Twisted Supersymmetric 5D Yang-Mills Theory and Contact Geometry,''
  JHEP {\bf 1205}, 125 (2012)
  [arXiv:1202.1956 [hep-th]].

\bibitem{Hama:2011ea}
  N.~Hama, K.~Hosomichi and S.~Lee,
  ``SUSY Gauge Theories on Squashed Three-Spheres,''
  JHEP {\bf 1105}, 014 (2011)
  [arXiv:1102.4716 [hep-th]].

\bibitem{Imamura:2011wg}
  Y.~Imamura and D.~Yokoyama,
  ``N=2 Supersymmetric Theories on Squashed Three-sphere,''
  Phys.\ Rev.\ D {\bf 85}, 025015 (2012)
  [arXiv:1109.4734 [hep-th]].

\bibitem{Nishioka:2013haa}
  T.~Nishioka and I.~Yaakov,
``Supersymmetric Renyi Entropy,''
  JHEP {\bf 1310}, 155 (2013)
  [arXiv:1306.2958 [hep-th]].

\bibitem{Dumitrescu:2012ha}
  T.~T.~Dumitrescu, G.~Festuccia and N.~Seiberg,
  ``Exploring Curved Superspace,''
  JHEP {\bf 1208}, 141 (2012)
  [arXiv:1205.1115 [hep-th]].

\bibitem{Closset:2012vg}
  C.~Closset, T.~T.~Dumitrescu, G.~Festuccia, Z.~Komargodski and N.~Seiberg,
  ``Contact Terms, Unitarity, and F-Maximization in Three-Dimensional Superconformal Theories,''
  JHEP {\bf 1210}, 053 (2012)
  [arXiv:1205.4142 [hep-th]].

\bibitem{Closset:2012vp}
  C.~Closset, T.~T.~Dumitrescu, G.~Festuccia, Z.~Komargodski and N.~Seiberg,
  ``Comments on Chern-Simons Contact Terms in Three Dimensions,''
  JHEP {\bf 1209}, 091 (2012)
  [arXiv:1206.5218 [hep-th]].

\bibitem{Closset:2012ru}
  C.~Closset, T.~T.~Dumitrescu, G.~Festuccia and Z.~Komargodski,
``Supersymmetric Field Theories on Three-Manifolds,''
  JHEP {\bf 1305}, 017 (2013)
  [arXiv:1212.3388 [hep-th]].

\bibitem{Festuccia:2011ws}
  G.~Festuccia and N.~Seiberg,
  ``Rigid Supersymmetric Theories in Curved Superspace,''
  JHEP {\bf 1106}, 114 (2011)
  [arXiv:1105.0689 [hep-th]].

\bibitem{Klare:2012gn}
  C.~Klare, A.~Tomasiello and A.~Zaffaroni,
  ``Supersymmetry on Curved Spaces and Holography,''
  JHEP {\bf 1208}, 061 (2012)
  [arXiv:1205.1062 [hep-th]].

\bibitem{Cassani:2012ri}
  D.~Cassani, C.~Klare, D.~Martelli, A.~Tomasiello and A.~Zaffaroni,
  ``Supersymmetry in Lorentzian Curved Spaces and Holography,''
  [arXiv:1207.2181 [hep-th]].

\bibitem{Hristov:2013spa}
  K.~Hristov, A.~Tomasiello and A.~Zaffaroni,
  ``Supersymmetry on Three-dimensional Lorentzian Curved Spaces and Black Hole Holography,''
  JHEP {\bf 1305}, 057 (2013)
  [arXiv:1302.5228 [hep-th]].

\bibitem{Fursaev:1995ef}
  D.~V.~Fursaev and S.~N.~Solodukhin,
  ``On the description of the Riemannian geometry in the presence of conical defects,''
  Phys.\ Rev.\ D {\bf 52}, 2133 (1995)
  [hep-th/9501127].

\bibitem{Klebanov:2011uf}
  I.~R.~Klebanov, S.~S.~Pufu, S.~Sachdev and B.~R.~Safdi,
  ``Renyi Entropies for Free Field Theories,''
  JHEP {\bf 1204}, 074 (2012)
  [arXiv:1111.6290 [hep-th]].

\bibitem{Alday:2013lba}
  L.~F.~Alday, D.~Martelli, P.~Richmond and J.~Sparks,
  ``Localization on Three-Manifolds,''
  [arXiv:1307.6848 [hep-th]].

\bibitem{Jafferis:2010un}
  D.~L.~Jafferis,
  ``The Exact Superconformal R-Symmetry Extremizes Z,''
  JHEP {\bf 1205}, 159 (2012)
  [arXiv:1012.3210 [hep-th]].

\bibitem{Hama:2010av}
  N.~Hama, K.~Hosomichi and S.~Lee,
  ``Notes on SUSY Gauge Theories on Three-Sphere,''
  JHEP {\bf 1103}, 127 (2011)
  [arXiv:1012.3512 [hep-th]].

\bibitem{Imamura:2011uw}
  Y.~Imamura,
  ``Relation between the 4d superconformal index and the $S^3$ partition function,''
  JHEP {\bf 1109}, 133 (2011)
  [arXiv:1104.4482 [hep-th]].

\bibitem{Nian:2013qwa}
  J.~Nian,
  ``Localization of Supersymmetric Chern-Simons-Matter Theory on a Squashed $S^3$ with $SU(2)\times U(1)$ Isometry,''
  [arXiv:1309.3266 [hep-th]].

\bibitem{Tanaka:2013dca}
  A.~Tanaka,
  ``Localization on round sphere revisited,''
  JHEP {\bf 1311}, 103 (2013)
  [arXiv:1309.4992 [hep-th]].

\bibitem{Closset:2013vra}
  C.~Closset, T.~T.~Dumitrescu, G.~Festuccia and Z.~Komargodski,
  ``The Geometry of Supersymmetric Partition Functions,''
  [arXiv:1309.5876 [hep-th]].

\bibitem{Martelli:2011fu}
  D.~Martelli, A.~Passias and J.~Sparks,
  ``The Gravity Dual of Supersymmetric Gauge Theories on a Squashed Three-sphere,''
  Nucl.\ Phys.\ B {\bf 864}, 840 (2012)
  [arXiv:1110.6400 [hep-th]].

\bibitem{Martelli:2011fw}
  D.~Martelli and J.~Sparks,
  ``The gravity dual of supersymmetric gauge theories on a biaxially squashed three-sphere,''
  Nucl.\ Phys.\ B {\bf 866}, 72 (2013)
  [arXiv:1111.6930 [hep-th]].

\bibitem{Martelli:2012sz}
  D.~Martelli, A.~Passias and J.~Sparks,
  ``The supersymmetric NUTs and bolts of holography,''
  Nucl.\ Phys.\ B {\bf 876}, 810 (2013)
  [arXiv:1212.4618 [hep-th]].

\bibitem{Martelli:2013aqa}
  D.~Martelli and A.~Passias,
  ``The gravity dual of supersymmetric gauge theories on a two-parameter deformed three-sphere,''
  Nucl.\ Phys.\ B {\bf 877}, 51 (2013)
  [arXiv:1306.3893 [hep-th]].

\bibitem{Casini:2011kv}
  H.~Casini, M.~Huerta and R.~C.~Myers,
  ``Towards a derivation of holographic entanglement entropy,''
  JHEP {\bf 1105}, 036 (2011)
  [arXiv:1102.0440 [hep-th]].

\bibitem{Hung:2011nu}
  L.~-Y.~Hung, R.~C.~Myers, M.~Smolkin and A.~Yale,
  ``Holographic Calculations of Renyi Entropy,''
  JHEP {\bf 1112}, 047 (2011)
  [arXiv:1110.1084 [hep-th]].

\bibitem{Belin:2013uta}
  A.~Belin, L.~-Y.~Hung, A.~Maloney, S.~Matsuura, R.~C.~Myers and T.~Sierens,
  ``Holographic Charged Renyi Entropies,''
  [arXiv:1310.4180 [hep-th]].

\bibitem{Casini:2010kt}
  H.~Casini and M.~Huerta,
  ``Entanglement Entropy for the n-sphere,''
  Phys.\ Lett.\ B {\bf 694}, 167 (2010)
  [arXiv:1007.1813 [hep-th]].

\bibitem
{mann} R.B.~Mann,
``Pair Production of Topological Anti-de Sitter Black Holes,"
Class. Quant. Grav. {\bf 14}, L109 (1997)
[gr-qc/9607071].

\bibitem
{brillet} D.R.~Brill, J.~Louko and P.~Peldan,
``Thermodynamics of (3+1)-dimensional Black Holes with Toroidal or
Higher Genus Horizons,"
Phys. Rev. {\bf D56}, 3600 (1997)
[gr-qc/9705012].

\bibitem
{vanzo}
L.~Vanzo,
  ``Black holes with unusual topology,''
  Phys.\ Rev.\ D {\bf 56}, 6475 (1997)
  [gr-qc/9705004].

\bibitem
{roberto}
R.~Emparan,
  ``AdS membranes wrapped on surfaces of arbitrary genus,''
  Phys.\ Lett.\ B {\bf 432}, 74 (1998)
  [hep-th/9804031].

\bibitem
{birm} D.~Birmingham,
  ``Topological black holes in Anti-de Sitter space,''
  Class.\ Quant.\ Grav.\  {\bf 16}, 1197 (1999)
  [hep-th/9808032].

\bibitem{Martelli:2011qj}
  D.~Martelli, J.~Sparks,
  ``The Large N Limit of Quiver Matrix Models and Sasaki-Einstein Manifolds,''
  Phys.\ Rev.\  {\bf D84}, 046008 (2011).
  [arXiv:1102.5289 [hep-th]].

\bibitem{Cheon:2011vi}
  S.~Cheon, H.~Kim and N.~Kim,
  ``Calculating the Partition Function of N=2 Gauge Theories on $S^3$ and AdS/CFT Correspondence,''
  JHEP {\bf 1105}, 134 (2011)
  [arXiv:1102.5565 [hep-th]].

\bibitem{Jafferis:2011zi}
  D.~L.~Jafferis, I.~R.~Klebanov, S.~S.~Pufu and B.~R.~Safdi,
  ``Towards the F-Theorem: N=2 Field Theories on the Three-Sphere,''
  JHEP {\bf 1106}, 102 (2011)
  [arXiv:1103.1181 [hep-th]].

\bibitem{Romans:1991nq}
  L.~J.~Romans,
``Supersymmetric, Cold and Lukewarm Black Holes in Cosmological Einstein-Maxwell Theory,''
  Nucl.\ Phys.\ B {\bf 383}, 395 (1992)
  [hep-th/9203018].

\bibitem{Freedman:1976aw}
  D.~Z.~Freedman and A.~K.~Das,
  ``Gauge Internal Symmetry in Extended Supergravity,''
  Nucl.\ Phys.\ B {\bf 120}, 221 (1977).

\bibitem{Caldarelli:1998hg}
  M.~M.~Caldarelli and D.~Klemm,
  ``Supersymmetry of Anti-de Sitter Black Holes,''
  Nucl.\ Phys.\ B {\bf 545}, 434 (1999)
  [hep-th/9808097].

\bibitem{Ryu:2006bv}
  S.~Ryu and T.~Takayanagi,
  ``Holographic derivation of entanglement entropy from AdS/CFT,''
  Phys.\ Rev.\ Lett.\  {\bf 96}, 181602 (2006)
  [hep-th/0603001].

\bibitem{Emparan:1999pm}
  R.~Emparan, C.~V.~Johnson and R.~C.~Myers,
``Surface Terms as Counterterms in the AdS / CFT Correspondence,''
  Phys.\ Rev.\ D {\bf 60}, 104001 (1999)
  [hep-th/9903238].

\bibitem{Lewkowycz:2013nqa}
  A.~Lewkowycz and J.~Maldacena,
``Generalized Gravitational Entropy,''
  JHEP {\bf 1308}, 090 (2013)
  [arXiv:1304.4926 [hep-th]].


\end{thebibliography}
\end{document}